\documentstyle[12pt,aasms4]{article}

\begin{document}

\title{The Influence of Nuclear Composition on the Electron Fraction in the 
Post-Core-Bounce
Supernova Environment}
\author{Gail C. McLaughlin and George M. Fuller}
\affil{Department of Physics, University of California, San Diego, La Jolla, 
CA 92093-0319}
\and 
\author{James R. Wilson}
\affil{Physical Science Directorate L-140, Lawrence Livermore National 
Laboratory, Livermore, CA 94550}
\authoremail{gail@bethe.ucsd.edu}

\begin{abstract}
We study the early evolution of the electron fraction (or, alternatively, the 
neutron-to-proton ratio)
in the region above the hot proto-neutron star formed after a supernova 
explosion.  We study the way in which
the electron fraction in this environment is set by a competition between 
lepton (electron, positron,
neutrino, and antineutrino) capture processes on free neutrons and protons and
 nuclei.  Our
calculations take explicit account of the effect of nuclear composition 
changes, such as formation of 
alpha particles (the \lq\lq alpha effect\rq\rq ) and the 
shifting 
of nuclear abundances in nuclear statistical equilibrium associated with 
cooling in near-adiabatic outflow.  
We take detailed account of the process of weak interaction freeze-out in 
conjunction with these
nuclear composition changes.
Our detailed treatment shows
 that the alpha effect can cause significant increases in the electron 
fraction, while neutrino
and antineutrino capture on heavy nuclei tends to have a buffering effect on 
this quantity.
We also examine the effect on weak rates and the electron fraction of 
fluctuations in time in the
neutrino and antineutrino energy spectra arising from hydrodynamic waves.
Our analysis is guided by the Mayle \& Wilson supernova
code numerical results for the neutrino energy spectra and density and 
velocity 
profiles.
\end{abstract}
\keywords{elementary particles - 
nuclear reactions, nucleosynthesis,
abundances - supernovae: general}
\section{Introduction}
\label{sec:introye}

In this paper we examine the early evolution of the electron fraction, $Y_e$, 
in the
post-core-bounce supernova environment.  The electron fraction is defined to 
be 
the net number of electrons per baryon:
\begin{equation}
\label{eq:defye}
Y_e = (n_{e^-} - n_{e^+}) / n_b   = 1 /(1 + N_n/ N_p);
\end{equation}
where $n_{e^-}$, $n_{e^+}$, and $n_b$ are the proper number densities of
electrons,
positrons, and baryons, respectively.  The latter expression in equation
(\ref{eq:defye}) follows from the condition of overall charge neutrality.  
Here,
$N_n$ and $N_p$ are the total proper neutron and proton densities, 
respectively,
so that $n_b = N_n + N_p$ and $N_n / N_p$ is the net neutron-to-proton ratio.

The electron fraction is a crucial determinant of nucleosynthesis produced from
neutrino-heated ejecta in models of post core-collapse supernovae (cf.,
\markcite{Woosley92} Woosley \&
Baron 1992; \markcite{Meyer92} 
Meyer et al. 1992; \markcite{Woosley94} Woosley et al. 1994; 
\markcite{Woosley92} Woosley \& Hoffman 1992;
\markcite{Qian93} Qian et al. 1993; \markcite{Qian96} Qian \& Woosley 1996; 
\markcite{Wiit93} Witti, Janka \& Takahashi 1993; \markcite{Hoffman96a} 
Hoffman et
al. 1996a; \markcite{Hoffman96b} Hoffman et al. 1996b).  We follow the 
terminology
of \markcite{Fuller92} Fuller, Mayle, Meyer, \& Wilson (1992),
\markcite{Fuller93} Fuller (1993) and \markcite{Qian95} Qian \& Fuller (1995)
 and
divide the post core bounce evolution of the outflowing material above the 
nascent
neutron star into two epochs: (1) the shock reheating or 
\lq\lq $p$-Process\rq\rq\ epoch at 
times post-core-bounce $t_{pb} \lesssim \, 1 \, {\rm s};$ and (2) 
the neutrino-driven wind or \lq\lq $r$-Process\rq\rq\ epoch at 
$t_{pb} \gtrsim \, 1 \, {\rm s}.$

We expect the shock reheating epoch to be characterized by chaotic outflow
(\markcite{Burrows95}Burrows, Hayes,  \& Fryxell, 1995; 
\markcite{Miller93} Miller, Wilson \& Mayle 1993; 
\markcite{Janka95}Janka \& M\"uller 1995; Herant, Benz, Hix, Fryer \& Colgate
1994; Janka \& M\"uller(1996)) and rapid heating by neutrino 
interactions of
material behind the shock.  Neutrino-heated ejecta originating from this epoch 
have
been suggested to give rise to the neutron number N=50 peak $r$-Process 
material
(\markcite{woosley92}Woosley \& Hoffman 1992; 
\markcite{Meyer92}Meyer et al. 1992;
\markcite{woosley94}Woosley et al. 1994) and possibly at least some of the 
light
$p$-Process nuclei such as $^{92}{\rm Mo}$ and $^{96}{\rm Ru}$ 
(\markcite{Fuller95}Fuller \& Meyer 1995; 
\markcite{Hoffman96a} Hoffman et al. 1996a).  However, in
the \markcite{Woosley94}Woosley et al. (1994) calculations (based on the Mayle
\& Wilson supernova results) 
 the N=50 $r$-Process nuclides originating in 
this epoch
are grossly overproduced relative to solar system abundances. Two fixes 
have been
proposed for the \lq\lq N=50 overproduction problem\rq\rq: (1) Fuller \& Meyer
(1995) invoke a modification of linear rapid outflow, a high neutrino 
fluence, and
the alpha effect to increase $Y_e$ and thereby reduce $N=50$ overproduction;
 and
(2) Hoffman et al. (1996a) show that as long as the electron fraction at this 
epoch
can be engineered to be $Y_e \gtrsim 0.484$, N=50 overproduction is avoided. 
 As a
bonus, both of these fixes concomitantly suggest that some of the light 
$p$-nuclei
will be synthesized.  Hoffman et al. (1996a) find that the 
light $p$-nuclei are produced in the correct
proportions so long as $0.484 \lesssim Y_e \lesssim 0.488$.
 This epoch is characterized by relatively low entropy per
baryon, $s/k \sim 40$, and relatively higher $Y_e$ compared to the conditions 
which
may obtain at $t_{pb} \gtrsim 1 \, {\rm s}$.  Note that the Hoffman et al. (1996a)
work implies that we may have to compute $Y_e$ to of order $1\%$ accuracy to
predict confidently the nucleosynthesis.  It could be that this will not be
necessary, as the hydrodynamic outflow is phased in just the right way that a
given mass element experiences the necessary $Y_e$ regime at the necessary
temperature.  Only future computations can address this issue.  As we will show, for
a given outflow model, predicting $Y_e$ histories to $1\%$ accuracy may be next to
impossible at this stage given our crude understanding of neutrino transport and
multidimensional hydrodynamic effects 
(cf. Herant, Benz, Hix, Fryer \& Colgate 1994; Janka \& M\"uller 1996).

By contrast, the later $r$-Process epoch is characterized by considerably 
higher
entropy, $s/k \approx 100-500$ (see \markcite{Qian} Qian \& Woosley 1996 and 
Meyer
et al. 1992), and possibly by a well ordered, near steady state, outflow 
resembling
a neutrino-driven wind (\markcite{Duncan86}Duncan, Shapiro, \& Wasserman 1986,
\markcite{Meyer92} Meyer et al. 1992, \markcite{Qian96} Qian \& Woosley 1996). 
 In
fact, \markcite{Woosley94} Woosley et al. (1994) have shown that the bulk of 
the
solar systems' $r$-Process material with nuclear mass $A \gtrsim 100$ could be
synthesized in this epoch. However, considerable controversy surrounds the
theoretical modeling of conditions in the \lq\lq hot bubble\rq\rq\ which 
forms in
this regime.  Though the \markcite{Woosley94} Woosley et al. (1994) 
calculations
yield a near perfect solar abundance distribution for the heavier $r$-Process
nuclides, they are based on the very high entropy ($s/k \sim 400$) conditions
obtained by the Wilson and Mayle supernova code.  Not only have such high 
entropies
been challenged (\markcite{Qian96}Qian \& Woosley 1996 find 
$s/k \lesssim 200$), but
even if the entropy were $s/k \gtrsim 300$, neutrino neutral current 
spallation of
alpha particles (\markcite{Meyer95}Meyer 1995) has been shown to result in a
drastic and fatal reduction in the neutron-to-seed ratio required for the
$r$-Process. Though models show the material in the hot bubble to be quite 
neutron
rich, $Y_e \approx 0.4$, \markcite{Hoffman1996b} Hoffman et al. (1996b) and 
\markcite{Luo} Meyer, Brown \&
Luo (1996) have demonstrated that far lower values of $Y_e$ are necessary to
obtain the requisite neutron-to-seed ratio for the $r$-Process if the entropy 
is
$s/k \lesssim 200$.  At entropies this low, the bad effects of neutrino
spallation of alpha particles
\markcite{Meyer95} (Meyer 1995) would be evaded.
Hoffman et al. (1996b) discuss the neutron-to-seed ratio and $Y_e$ issues 
related
to this epoch, while \markcite{Fuller96} Fuller, Qian, \& Wilson (1996) and
\markcite{Caldwell96} Caldwell, Fuller \& Qian (1996) discuss neutrino flavor
mixing schemes which could give lower $Y_e$ and hence help the $r$-Process.

In this paper we perform a complimentary study of the evolution of the electron
fraction which concentrates on the effects of nuclear composition changes.  We
focus in particular on the early time, 
\lq\lq low\rq\rq\ entropy environment of the shock
reheating epoch.  In what follows we concentrate on the weak interaction balance
essentially in a single outflowing mass element (i. e., one-dimensional outflow). 
We employ outflow results from the Mayle \& Wilson supernova code to illustrate
various effects bearing on $Y_e$.  It should be kept in mind that we are not 
{\it predicting} $Y_e$, as the Mayle and Wilson results may not be representative
of the true picture of supernova evolution.  For example, multidimensional
hydrodynamic effects could effectively cause different mass elements to have
different time/thermodynamic histories.  Nevertheless, we choose the simplest case
(1D outflow) to elucidate the physics.   

In section \ref{sec:over}, we present an overview of all of 
the variables
which affect the calculation of the electron fraction.  We show explicitly 
how the various charged current lepton capture processes
 are important.  In section \ref{sec:nucap}, we discuss
salient aspects of the electron neutrino and antineutrino capture rates on free
nucleons.  We explore the difference between using the Mayle and
Wilson transport
calculation-derived neutrino energy spectra as an example
 and approximate blackbody spectra.  We also discuss
the effect on the electron fraction 
of hydrodynamic wave-induced fluctuations in the neutrino energy 
spectra.  In section \ref{sec:elcap}, we assess the role of electron and
 positron capture 
processes on
$Y_e$.  In section \ref{sec:alpha}, we examine the \lq\lq alpha effect\rq\rq\ or the 
tendency of
the formation of alpha particles to raise the electron fraction.  We discuss 
the
equilibrium and nonequilibrium nature of weak interaction freeze-out in 
section \ref{sec:noneq},
with particular attention to the role of nuclear composition changes and the 
role
of neutrino capture on heavy nuclei.  We give conclusions in section \ref{sec:conclusion}.  

\section{Overview}
\label{sec:over}

The electron fraction, which is defined in equation (\ref{eq:defye}), can be 
written in the
following way,
\begin{equation}
\label{eq:defye2}
Y_e = \sum_{i} (Z_i/A_i) X_i = \sum_{i} Z_i Y_i.
\end{equation}
Here we assume overall plasma charge neutrality, so that the sum in equation 
(\ref{eq:defye2})
runs over all nuclear species $i$ with charge $Z_i$, nuclear mass number 
$A_i$, 
mass fraction
$X_i$, and number abundance relative to baryons $Y_i = X_i/ A_i$.  Typically, 
the material is very
hot near the surface of the neutron star, where essentially
 all of the baryons are in free 
nucleons.  As the
material flows away from the neutron star, it cools and alpha particles  
begin to form.  As it flows farther out and cools further, heavier nuclei near
the iron peak begin to form.  With this rough evolution of abundances with
radius and time in mind, we can rewrite equation 
(\ref{eq:defye2}) as,
\begin{equation}
 Y_e = X_p + X_\alpha/2 + \sum_{h} (Z_h/A_h) X_h,
\end{equation}
where $X_p$ is the mass fraction of free protons, and $X_\alpha$ is the mass
fraction of alpha particles,  and the summation runs over all nuclear species
 $h$
heavier than alpha particles.  In the conditions common in neutrino-heated
outflow, \lq\lq free\rq\rq\ (not inside nuclei) neutrons and 
protons,
alpha particles, and a few iron peak nuclei typically 
account for most of the baryons.

The charged current weak interactions alter the electron fraction by converting
neutrons into protons and {\it vice versa}. Most important in the region
above the neutron star are neutrino and 
antineutrino capture on free nucleons and the associated reverse processes:
\begin{equation}
\label{eq:freenu}
\nu_{e} + {\rm n} \leftrightarrow {\rm p} + e^{-};
\end{equation}
\begin{equation}
\label{eq:freenu2}
\bar{\nu}_{e} + {\rm p} \leftrightarrow {\rm n} + e^{+}.
\end{equation}
However, the processes of electron neutrino and antineutrino capture on heavy
nuclei can sometimes be important in determining the overall neutron-to-proton
balance (\markcite{Fuller95}Fuller \& Meyer 1995;
\markcite{Mclaughlin95}McLaughlin \& Fuller 1995):
\begin{equation}
\label{eq:heavynu}
{\nu}_e + {\rm A(Z,N)} \rightarrow  {\rm A(Z+1,N-1)} +  
e^{-};
\end{equation}
\begin{equation}
\label{eq:heavynu2}
{\bar{\nu}}_e + {\rm A(Z,N)} \rightarrow  {\rm A(Z-1,N+1)} +  e^{+}.
\end{equation}
In these expressions, A, Z, and N are the total number of nucleons, proton 
number
and neutron number of the nucleus, respectively.  The reverse rates of electron
and positron capture on heavy nuclei are generally negligible for the 
conditions
in which these nuclei form.  The ratio of neutrons-to-protons and $Y_e$ in
neutrino-heated
material flowing away from the neutron star is set by a competition between the
rates of the processes in equations (\ref{eq:freenu}) and (\ref{eq:freenu2}) 
and
(\ref{eq:heavynu}) and (\ref{eq:heavynu2}) and the overall material 
expansion rate (or
outflow rate).  In fact, it has been shown that where the rates of these
processes are rapid compared to the outflow rate, a characteristic weak 
steady state or \lq\lq chemical equilibrium\rq\rq\ obtains 
(\markcite{Qian93}Qian et al. 1993; \markcite{Qian93a} Qian 1993).  The weak
freeze-out radius is defined to be where the rate of 
${\rm n} \rightleftharpoons {\rm p}$ interconversion as set by the rates of 
the processes in equations 
(\ref{eq:freenu}), 
(\ref{eq:freenu2}), (\ref{eq:heavynu}) and (\ref{eq:heavynu2})
falls below the material outflow rate.  
Though the details are complicated, $Y_e$
at small radius is set principally by the processes in equations 
(\ref{eq:freenu}) and (\ref{eq:freenu2}); whereas, at larger radius and at 
later
times, the processes in (\ref{eq:heavynu}) and (\ref{eq:heavynu2}) can also 
make a contribution.

The rate of change of the electron fraction of an outflowing mass element
resulting from weak interactions may be written as follows 
(this is a generalization of the treatment in Qian et al. 1993):
\begin{equation}
\label{eq:yedot}
{dY_e \over dt} = v(r) {dY_e \over dr}
 = (\lambda_{\nu_e} + \lambda_{e^+}) X_n - (\lambda_{\bar{\nu}_e}
+ \lambda_{e^-}) X_p + \sum_h \left({X_h \over A_h}\right) (\lambda_{h\nu_e} + 
\lambda_{he^+} - \lambda_{h\bar{\nu}_e} - \lambda_{he^-} ),
\end{equation}
where $v(r)$ is the radial velocity field above the neutron star, and $t$ is 
a time development parameter.
In equation (\ref{eq:yedot}), the sum on $h$ indicates a sum over all heavy 
nuclei.  Here $X_{n}$ is the mass fraction of free neutrons,
while $X_h$ is the appropriate mass fraction of a heavy nucleus
with mass number $A_h$.  We
define $\lambda_{\nu_e}$ and $\lambda_{\bar{\nu}_e}$ to be the forward rates
 of the processes in equations (\ref{eq:freenu}) and (\ref{eq:freenu2}),
respectively.  Similarly, $ \lambda_{e^-}$ and $\lambda_{e^+}$ are the 
reverse rates of these processes, respectively.  Finally, 
$\lambda_{h\nu_e} (\lambda_{he^+}) $ and 
$\lambda_{h\bar{\nu}_e} (\lambda_{he^-})$ represent the electron 
neutrino (positron) and antineutrino 
(electron) capture rates, respectively, on the nucleus with 
index h.  Note that equation (8) contains {\it all} the effects which 
bear on $Y_e$
and the neutron-to-proton ratio in a given mass element.  Hydrodynamic 
motion can
influence this weak balance via the position of the weak freeze-out radius 
(Qian et
al. 1993) which is determined largely by the velocity field $v(r)$ (more 
properly, in
multidimensional hydrodynamic regimes we should write $v(r,\theta,\phi,t)$).

Near the neutron star, before the iron peak nuclei form, the last term
in equation (8) is zero.  
The neutrino and antineutrino capture reactions on free nucleons 
are usually the fastest of the nuclear charged current weak interaction 
processes. 
The exception occurs in the region very 
close to the neutron star, where electron and positron
capture on free nucleons become comparable to the corresponding neutrino 
capture rates.  In 
the region near the neutron star where the electrons are
relativistically degenerate, $\lambda_{e^-}$  can be large and can 
essentially
set $Y_e$ to quite low values.  By contrast, at late times and/or large 
radius,
the free nucleons become absorbed into nuclei, and the last term in equation
(\ref{eq:yedot}) becomes larger than or competitive with the other two 
terms (in this environment, $\lambda_{he^+} $ and $\lambda_{he^-}$ are always
small compared to $\lambda_{h\bar{\nu}_e}$ and/or $\lambda_{h\nu_e}$). 
Here we include no term which is proportional to $X_\alpha$, 
since charged current
capture rates on alpha particles are too small to influence the electron
fraction in the relevant conditions.

We can give a formal solution to equation (\ref{eq:yedot}) in the limit where
there are free nucleons and alpha particles only, and no heavy nuclei:
\begin{displaymath}
Y_e = 
\lbrace{Y_{e_i} - {[X_\alpha/2 + \lambda_n /(\lambda_n +
 \lambda_p)]}_i }\rbrace \exp{\left[ -\Large\int_{t_i}^{t_f} 
(\lambda_n + \lambda_p) dt \right]}
\end{displaymath}
\begin{displaymath}
 +{[X_\alpha/2 + \lambda_n/(\lambda_p + \lambda_n)]}_f 
\end{displaymath}
\begin{equation}
\label{eq:soldiff}
\quad \quad \quad \quad -\int_{t_i}^{t_f} {d\over dt} {\left[ X_{\alpha} /2 
+   \lambda_n/(\lambda_p + \lambda_n)\right]} 
    \exp{\left[-\int_{t}^{t_f} (\lambda_n + \lambda_p) dt^\prime \right]} 
dt. 
\end{equation}

In this expression the initial conditions are denoted by the subscript 
$i$ and the final
conditions are denoted by the subscript $f$. Here, 
$\lambda_n \equiv \lambda_{e^+} +
\lambda_{\nu_e}$ and 
$\lambda_p \equiv \lambda_{e^{-}} + \lambda_{\bar{\nu}_e}$ are
the total neutron and proton
 destruction rates, respectively, resulting from weak
interaction processes.  When the system is in weak equilibrium, the rate of
change  of the conditions with time or radius
 in the plasma will be slow compared
with the magnitude of the weak rates $\lambda_n$ and $\lambda_p$.  The first 
term
in equation (\ref{eq:soldiff}) will be negligible when weak equilibrium 
obtains. 
The second term in this equation will be the value of the electron fraction 
as the
weak equilibrium limit is approached.  
The exponential factor in the integrand of the last
term is very small except for $t$ near $t_f$.  This factor will not be small 
for
integration intervals, $t - t_f = \delta t \sim 1/(\lambda_n + \lambda_p).$
As long as the product of the time derivatives of $X_{\alpha}$ and 
$\lambda_n/(\lambda_p + \lambda_n)$ with $\delta t$ is
small, the last term in equation (\ref{eq:soldiff}) will also be small.  This 
condition
is usually satisfied where weak equilibrium obtains and where the alpha 
particle mass
fraction is only slowly varying with time.  If the system is not in weak
equilibrium, and/or the alpha particle mass fraction is changing significantly
with time, then in principle, all of the terms in equation (\ref{eq:soldiff})
may be necessary for calculating the electron fraction.

If weak equilibrium is established, alpha particles are absent,
 and the electron and positron capture rates, 
$ \lambda_{e^-}$ and $\lambda_{e^+}$,  
are both zero, then the electron fraction will be,
\begin{equation}
Y_e = 1/(1 + \lambda_{\bar{\nu}_e}/\lambda_{\nu_e}) 
\equiv Y_{e_0}.
\end{equation}
This will be a reasonable approximation to the true electron faction 
at about $t_{pb} \approx 1 \, {\rm
s}$ in the region just below where alpha particles form (\markcite{Qian93a}Qian
et al. 1993). We shall use the quantity $Y_{e_0}$ as a
first estimate of the electron fraction.  We discuss computational estimates 
of $Y_{e_0}$ in detail in section \ref{sec:nucap}.

Although $Y_{e_0}$ provides a good first estimate of the electron fraction, the
electron and positron capture rates on free nucleons 
can also have some influence.  In fact, at a
plasma temperature of $\sim 2 \, {\rm  MeV}$, the electron and positron capture
rates can make a significant contribution in the determination of the 
electron fraction.  If the system is in weak equilibrium and alpha particles 
are
absent, then the electron fraction in this regime of high plasma temperature is
more accurately given by,
\begin{equation}
\label{eq:yeeqel}
Y_e = 1/[1 + (\lambda_{\bar{\nu}_e}+\lambda_{e^-})
/(\lambda_{e^+} +\lambda_{\nu_e})] = (1 + \lambda_p / \lambda_n )^{-1}.
\end{equation}
This can be seen from equation (\ref{eq:soldiff}) above.  At such high plasma
temperature, $X_\alpha \approx 0$ usually will be a good approximation for 
the
relevant entropies in the supernova models we consider.  The effect of 
electron
and positron capture on $Y_e$ will be discussed in section \ref{sec:elcap}.  
We generalize
this discussion to include the $X_\alpha \neq 0$ case in section \ref{sec:alpha}.

\markcite {Fuller95}
Fuller \& Meyer (1995) first pointed out that the mass fraction of alpha
particles can influence $Y_e$.  This effect can be readily gleaned from 
equation
(\ref{eq:soldiff}).
When the alpha mass fraction rises (plasma temperature $\sim$ 0.5 MeV), 
free 
nucleons are typically absorbed into alpha
particles.  Each alpha particle removes an equal number of free protons (2) 
and free neutrons (2) when it forms.  If the outflowing material in the 
supernova is
neutron-rich before the alpha particle formation, then as the alpha particle 
mass
fraction increases the residual free nucleon gas will tend to be enriched in
neutrons.  Since the charged current capture
rates on alpha particles are negligible in this situation, 
the only interactions that change the electron
fraction are lepton 
captures on free nucleons.  However, since alpha particle formation
has left mostly free neutrons, the neutrino and positron capture
 interactions will have the effect of
turning some of these \lq\lq leftover\rq\rq\ free neutrons into protons.
In turn, this will cause the total electron fraction to rise.
We follow \markcite{Fuller95}Fuller \& Meyer (1995) and 
call this increase in $Y_e$ due to alpha particle formation and weak
interactions the \lq\lq alpha 
effect.\rq\rq\  We give a detailed treatment of this effect in section 
\ref{sec:alpha}.

Since the material may not be in weak equilibrium at the time of nuclear 
reaction
rate freeze-out and nucleosynthesis, it is necessary to examine the effect on
the
electron fraction of slow freeze out from weak equilibrium.  This is done in
section \ref{sec:noneq}, where we present the results of a nonequilibrium 
calculation of $Y_e$. 
The influence  on $Y_e$ 
 of neutrino captures on heavy nuclei will also be
discussed in section \ref{sec:noneq}.  
As is evident from equation (\ref{eq:yedot}), these processes
in principle can impact the final value of the electron fraction.  Although it
will turn out that the material is no longer in weak equilibrium at the time of
formation of iron peak nuclei in the models we consider, the neutrino 
captures on
heavy nuclei (plasma temperature $\sim$ 0.3 MeV)
still have some influence on $Y_e$.

\section{Neutrino and Antineutrino Capture Rates on Free Nucleons}
\label{sec:nucap}

In this section we compare the relative magnitude and effects on $Y_e$ of
$\lambda_{\nu_e}$ and $\lambda_{\bar{\nu}_e}$. For the calculation of these
rates we follow the prescription of
\markcite{Fuller95}Fuller \& Meyer (1995; 1996) and 
\markcite{McLaughlin95}McLaughlin \& Fuller (1995; 1996).  In computing these 
rates, we
first assume a blackbody distribution  for the neutrino spectrum, 
which we normalize by the neutrino luminosity  (see 
\markcite{Qian95}Qian \& Fuller 1995).  
In Figure \ref{fig:compblack} we depict a zero
chemical potential blackbody neutrino $(\nu_e)$ energy spectrum 
(curve without
circles) with average neutrino energy and normalizing luminosity taken from 
the
numerical results of the Mayle and Wilson supernova calculations at $t_{pb}
\approx 0.575 \, {\rm s}^{-1}$.  Also shown in this figure is the actual 
neutrino
energy spectrum (circles) at large radius at $t_{pb}
\approx 0.575 \, {\rm s}^{-1}$ as given by the detailed transport 
calculations of Mayle and Wilson.  Of course, the {\it actual} neutrino
energy spectra may differ considerably between different models which employ
different transport calculations.  We here use the Mayle and Wilson results
only as an example.

In our calculations of $\lambda_{\nu_e}$ and $\lambda_{\bar{\nu}_e}$ we 
generally
assume that the material is far from the neutron star, so that the distance
dependence of the rates is simply  $\propto
r_7^{-2}$, where $r_7$ is the distance from the center of the
neutron star in units of $10^7
{\rm cm}$.  We 
further assume that
the final state electrons are very relativistic, so that the rate may be
expressed with Fermi integrals.  In evaluating the Fermi intergals for the 
antineutrino
capture rates, we use the approximation
$  {( m_p - m_n - m_e)} / T_{\bar{\nu}} \ll 0$, where $m_n$ is the mass of 
the
neutron, $m_p$ is the mass of the proton, $m_e$ is the mass of the electron,
and $T_{\bar{\nu}}$ is the temperature of the 
antineutrino distribution function 
(we
assume zero chemical potential).  The capture rates are then approximately,
\begin{equation}
\label{eq:nuapp} 
 \lambda_{\nu_e} 
\approx (0.1945 \, {\rm s}^{-1}) 
{\left(L_\nu \over 10^{51} {\rm erg \, s^{-1}} \right)}
{\left( T_\nu \over {\rm  MeV} \right)} \left(1 \over r^2_7 \right) C_1;
\end{equation}
\begin{equation}
\label{eq:nuapp2}
\lambda_{\bar{\nu}_e} \approx (0.2000 \, {\rm s}^{-1}) 
\exp{\left(-1.804 \, {\rm MeV} \over T_\nu \right)}
{\left(L_{\bar{\nu}} \over 10^{51} {\rm erg \, s^{-1}} \right)} 
{\left( T_{\bar{\nu}} 
\over {\rm MeV} \right)} {\left(1 \over r^2_7 \right)} C_2.
\end{equation}
In this expression we have approximated the initial and final state lepton
kinematics as completely relativistic, and we have neglected final state lepton
blocking \markcite{McLaughlin95}(McLaughlin \& Fuller 1995).
The terms $C_1$ and $C_2$ are defined in the following way:
\begin{equation}
C_1 \approx  1 + (0.6283 \, {\rm MeV})/T_\nu + 
(0.1292  \, {\rm MeV}^2)/T_\nu^2;
\end{equation}
\begin{equation}
C_2 \approx 1+ (1.158  \, {\rm MeV})/T_{\bar{\nu}} + 
(0.600  \, {\rm MeV}^2)/T_{\bar{\nu}}^2 +
(0.1778  \, {\rm MeV}^3)/T_{\bar{\nu}}^3.
\end{equation}
These estimates demonstrate the dependence of the rates on neutrino
(antineutrino) luminosity, $L_{\nu({\bar{\nu}})}$, the
temperature, $ T_{\nu ({\bar{\nu}})}$, and distance from the neutron star 
center, $r_7$.
The rates $\lambda_{\bar{\nu}_e}$ and $\lambda_{\nu_e}$ are approximately
 proportional to the antineutrino and neutrino temperature, respectively, when
these temperatures are large.  However, at lower neutrino and antineutrino
temperature they have
a more complicated temperature dependence through the terms $C_1$ and $C_2$ and
the exponential term in equation (\ref{eq:nuapp2}). 
The exponential term in the antineutrino capture rate has 
its origin in the energy threshold, $m_n - m_p + m_e$. Of course, there is 
not a 
corresponding term in the neutrino capture rate, since neutrino capture on 
free 
neutrons has no energy threshold.   We
estimate the temperature of the blackbody distribution to be 
$\approx \bar{E}/3.15$,
where $\bar{E}$ is the average energy of the neutrino or antineutrino 
spectrum as
appropriate.  At $t_{pb} \approx 0.500 \, {\rm s}$, the neutrino temperature 
is $T_\nu \approx 3.85 \, {\rm MeV}$ and
the antineutrino temperature is $T_{\bar{\nu}} \approx 4.93 \, {\rm MeV}$ in 
the
Mayle and Wilson numerical calculations.  

The ratio of the antineutrino and
neutrino capture rates on free nucleons is
\begin{equation}
\label{eq:nuratio}
{\lambda_{\bar{\nu}_e} \over \lambda_{\nu_e}} \approx 
1.029 \left( L_{\bar{\nu}} \over 
L_\nu \right) \left(T_{\bar{\nu}} \over T_\nu \right) 
\exp{[(-1.804 \, {\rm MeV})/T_{\bar{\nu}}]} J_1,
\end{equation}
where the term $J_1$ is defined as,
\begin{equation}
J_1  \equiv {\left( C_2 \over C_1 \right)} \approx 1 + (1.158 \, {\rm
MeV)} /T_{\bar{\nu}} 
- (0.6283 \, {\rm MeV})/T_\nu.
\end{equation}
Since the neutrino and antineutrino temperatures are relatively high, we have
ignored terms of higher order in these temperatures in $J_1$.  Because
 of
the neutron excess in the neutron star, $T_{\bar{\nu}_e}$ will be generally
larger than $T_{\nu_e}$ everywhere above the neutrino sphere (we ignore 
possible
neutrino flavor mixing effects - see \markcite{Qian93}Qian et al. 1993, and 
\markcite{\Qian95}Qian \& Fuller 1995). Also, at this epoch, the total
 energy luminosity of the
neutrinos is about 10\% smaller than that of the antineutrinos in the Mayle
and Wilson supernova results.
In order for the
antineutrino capture rate to dominate over the neutrino capture rate, the
antineutrino temperature and luminosity must be sufficiently greater than the
corresponding neutrino temperature and luminosity
to overcome the effect of the energy threshold.
Since the neutrino and antineutrino capture rates have the same dependence on
distance from the neutron star, the relative importance of the rates does not
change explicitly with radius in this formulation. 

The $\nu_e$ and $\bar{\nu}_e$ 
temperatures and luminosities change with time, and therefore the neutrino
and antineutrino capture rates are changing implicitly 
as the material flows away from the neutron star.  The
ratio of our approximate
 antineutrino capture rate on protons 
to our approximate neutrino capture rate on neutrons has been
plotted as a function of time (or epoch) 
in Figure \ref{fig:ratios1}.  The antineutrino capture rate is always larger 
that the
neutrino capture rate in the indicated time interval,
 and this difference is increasing with time.  The curve in 
Figure \ref{fig:ratios1}
has been smoothed into a straight line for illustrative purposes.  Variations
of this ratio from a strictly increasing function will be discussed further 
below.
   
Since the  ratio 
$\lambda_{\bar{\nu}_e} / \lambda_{\nu_e}$ is important for the calculation of 
the electron fraction,
it is better to use the numerically computed energy spectrum to calculate the
neutrino and antineutrino capture rates.  
An example of such a numerically calculated energy spectrum
 taken from the results of
the Wilson and Mayle code is shown in Figure \ref{fig:compblack}.  The circles 
in Figure \ref{fig:compblack} show the points
produced by the detailed neutrino transport computations of this code.
In order to calculate the capture rates from these numerical points
it is first necessary to interpolate to find a value of the energy occupation
probability for all neutrino
energies.  Given such an interpolation scheme, the capture rates can be
calculated with the expressions, 
\begin{equation}
\label{eq:nuexact}
\lambda \approx {1 - (1 - (R_{\nu (\bar{\nu})}/r)^2)^{1/2} \over 
1 - (1 - (R_{\nu (\bar{\nu})}/r_0)^2)^{1/2}}
 \int_{E_{TH}}^{\infty} \sigma (E) f_{\nu (\bar{\nu})}(E) dE.
\end{equation}
Here, $f_{\nu (\bar{\nu})}(E$) is the interpolated value for the number flux 
of 
(anti)neutrinos at energy $E$.  
The (anti)neutrino sphere radius is given by $R_{\nu (\bar{\nu})}$, and 
r is the distance from
the neutron star center at which the rate is to be calculated. The distance 
from 
the neutron star center 
at which the neutrino spectrum has been evaluated is $r_0$.   In the
case of the Wilson and Mayle code, this distance is $r_0 = 3 \times 10^7 {\rm
cm}$. Assuming relativistic lepton kinematics, and no final state lepton 
blocking, the cross section at energy $E$ is given by
\begin{equation}
\label{eq:nucross}
\sigma (E)  \approx (9.542 \times 10^{-44} {\rm cm}^2) \left({E + Q \over
{\rm MeV}} \right)^2,
\end{equation}
where $Q$ is the approximate nuclear $Q$-value (see equation 4 in 
McLaughlin \&
Fuller 1995). The $Q$-value for
 neutrino capture on neutrons is 
$Q \approx 1.293 \, {\rm MeV}$, while for antineutrino
capture on protons it is 
$Q \approx -1.293 \, {\rm MeV}$.  The energy threshold is
$E_{TH} = 0$ for neutrino capture on neutrons while it is $E_{TH} = (m_n -
m_p + m_e) \approx 1.804 \, {\rm MeV}$ for antineutrino capture on protons.  

Table \ref{tab:nccc} illustrates the discrepancies in 
capture rates and electron fractions calculated by different methods.  
In the first column of this table, the method of calculating the neutrino
or antineutrino spectrum is given.  In the second and third columns, the
corresponding neutrino and antineutrino capture rates on free neutrons and
protons, respectively, are given.  In the fourth column, the quantity,
$Y_{e_0} = 1/(1 + \lambda_{\bar{\nu}_e} / \lambda_{\nu_e})$ is tabulated.
All rates are calculated at a distance from the neutron star center of $r =
4.793 \times 10^6 \,  {\rm cm}.$  
All quantities are calculated at time $t_{pb}
= 0.575 \, {\rm s}$ in the Mayle and Wilson calculations.  The ${\nu_e}
({\bar{\nu}_e})$ temperatures employed for the blackbody spectra in the
first row are $T_{\nu_e} = 3.724 \, {\rm MeV}$ and $T_{\bar{\nu}_e} = 4.835
\, {\rm MeV}.$

The first row of Table \ref{tab:nccc} 
shows the results of the calculations with zero chemical potential
 blackbody neutrino and antineutrino energy spectra.  Such a blackbody 
energy distribution tends to overestimate the (anti)neutrino capture rate.  
This
is due to depletion in the high energy tail of the actual transport 
calculation-derived neutrino energy spectrum relative to the zero chemical 
potential
blackbody approximation.  This depletion is caused by (anti)neutrino
absorption and scattering above the neutrino sphere.  The blackbody spectrum
tends to underestimate the electron fraction, as shown in the column labeled
$Y_{e_0}$.

The second row of Table \ref{tab:nccc} gives the resulting 
capture rates and electron fraction when cubic spline interpolation is used to
generate the segments of curve between the points on the 
numerically-derived (anti)neutrino energy spectra, and
equations (\ref{eq:nuexact}) and 
(\ref{eq:nucross}) are used to calculate the capture rates.
In order to check the suitability of this method for our purposes,
we tried another interpolation scheme to estimate the capture rates.
We fit curves between the numerically-derived spectrum 
points with pieces of blackbody distribution.  The Mayle and Wilson spectrum
points themselves are derived from the numerical transport calculations and 
are
produced with an interpolation procedure in some ways similar to that employed
here.

Each such blackbody segment had a
different temperature.  As the neutrinos leave the neutron star,
different energy neutrinos decouple at different radii and temperature.  This
motivated our fitting of the neutrino spectrum by pieces of different thermal
distributions.  The results are shown in the third row of Table 
\ref{tab:nccc}.
The difference between
the approximate electron fraction $Y_{e_0}$ obtained in this last case and in 
the
case of cubic spline interpolation was about 0.3\%.  
  The difference in $Y_{e_0}$  between the case with 
cubic spline
interpolation and the case with a single zero chemical potential
 blackbody distribution was about 4\%.   Since it may be necessary
to obtain the value of the electron fraction to within 1\% (Hoffman et al. 
1996a;
Qian \& Woosley 1996),  
cubic spline interpolation gives an adequate result.  This
is the method that we have employed in our calculations of the neutrino and
antineutrino capture rates throughout the remainder of this paper.

Figure \ref{fig:ye0vst} shows the value of the parameter $Y_{e_0}$, at 
different
times, calculated using cubic spline interpolation for the neutrino and
antineutrino energy spectra taken from a particular run of the Wilson \& Mayle
supernova code (crosses). 
There is significant variation in this quantity during the time increment
$(\Delta t \approx 0.025 \, {\rm s^{-1}})$ between typical snapshots of the
neutrino and antineutrino energy spectra at $t_{pb} \lesssim 0.575 \, {
\rm s}$.
In the hot bubble above the neutrino sphere, hydrodynamical waves are 
propagating back and forth. Spherical convergence may amplify these waves near
the neutrino sphere.  Multi-dimensional non-spherical geometries may
complicate this picture considerably (Burrows et al. 1995; Janka \& M\"uller
1995).
The variations evident in Figure \ref{fig:ye0vst} may be
due to these physical effects.  The value of $Y_{e_0}$
obtained with a similar procedure, but employing a different Mayle and Wilson
numerical model for this epoch, suggests smaller variations.
In any case, since we may 
need to compute $Y_e$ to of order $\sim 1 \%$ accuracy 
for nucleosynthesis purposes (Hoffman et al. 1996a), it is important 
to be aware of
this issue. 
Since this variation in $Y_{e_0}$
 is relatively large $\sim 3\%$, it will be necessary
 to resolve this and other issues before the
electron fraction can be reliably calculated to accuracies of
approximately $\sim 1\%$ during this epoch.  It is obvious from these
considerations that other supernova outflow models, employing different
(anti)neutrino energy spectra and/or different hydrodynamic conditions 
may well
give a $Y_e$ very different from that computed here.  Our results are 
meant to be
only illustrative of the physics required to estimate $Y_e$ to $1\%$ 
for any model.  

For illustrative
purposes we have chosen to draw a straight line through the time period
$0.575 \, {\rm s} - 0.700 \, {\rm s}$ (during which the point-to-point 
variations in $Y_{e_0}$ are small) yielding the linear relation,
\begin{equation}
Y_{e_0} \approx 0.4606 - 0.7120(t - 0.5750).
\end{equation}
This line, which is shown in Figure \ref{fig:ye0vst}, is only meant to 
demonstrate the effect of the increasing
disparity between the neutrino and antineutrino energy spectra with time on
$Y_{e_0}$.

The neutrino and antineutrino energy spectral data
  that we use here comes from snapshots taken from the Mayle \& Wilson 
output at various times post-bounce.   At each time, these spectra have has 
been evaluated at a radius of $3 \times 10^7 {\rm cm}$.  
However, the neutrino and antineutrino energy spectra
 will change at different points in the region above the neutron star
due to weak interaction processes in  the plasma.  For the situation we are
considering, we are far from the neutrino sphere, and close to the radius 
$r_0$  of spectral quantity 
evaluation.  However, it is clear that the effects of neutrino emission,
absorption, and scattering must also be 
included for a precise
determination of $Y_{e_0}$ in the region near the neutron star. 
We believe that these effects are
smaller than the observed oscillation in the quantity $Y_{e_0}$ apparently
caused by local variations from hydrodynamic waves.
\section{Electron and Positron Capture Rates on Free Nucleons}
\label{sec:elcap}

The processes of electron and positron capture on free nucleons are given in
equations (\ref{eq:freenu}) and (\ref{eq:freenu2}).
These processes are important in regions close to the
neutron star.  They play a role in determining the position
 of the gain radius.  Neutrino interactions cause a net positive heating of
the region above the gain radius (Bethe \& Wilson 1985). 
 At an early epoch $(t_{pb} \lesssim 1s)$,
 the weak freeze out radius, or distance from the neutron star center at
which the characteristic 
rate of weak interactions becomes smaller than the rate of material outflow
in the supernova, occurs
sufficiently beyond the gain radius so that electron and positron capture rates
on free nucleons
have fallen considerably below the corresponding 
neutrino and antineutrino capture rates.
At this distance, and at all points further away from the neutron star, the
electron (positron) capture rates are always much smaller than the 
(anti)neutrino capture rates on free nucleons.

Here we give estimates of the electron and positron capture rates 
on free nucleons, analogous to the
estimates of neutrino and antineutrino capture rates presented in
section \ref{sec:nucap} (equations \ref{eq:nuapp} and \ref{eq:nuapp2}).  
We employ Fermi-Dirac distributions for 
the electrons and positrons, since these particles are well approximated as
being in thermal
equilibrium with the plasma.
This is a much different situation than the 
case
of neutrinos, since the neutrinos decouple from the plasma very close to the
neutron star.  In producing these estimates,
 we have assumed that the electrons are relativistic, so that the
appropriate phase space factors
 may be reduced to Fermi integrals (\markcite{Fuller85}Fuller, 
Fowler, \& Newman 1985; \markcite{Fuller82ab}1982ab; 
\markcite{Fuller80}1980; \markcite{Fuller82}Fuller
1982).  Although the electrons and
positrons are not completely 
relativistic throughout the entire period of interest for 
nucleosynthesis, their capture
rates on free nucleons
are only important at fairly high temperatures $(T_e > 1 {\rm MeV})$ 
and relatively small distances
from the neutron star.  Therefore, this approximation is sufficient to
demonstrate the general effect of
these capture rates on the electron fraction.  
The estimates of the rates are given by,
\begin{equation}
\label{eq:e-rate}
\lambda_{e^-} \approx (1.578 \times 10^{-2} \,{\rm s}^{-1})
\left({T_e \over m_e c^2} 
\right)^5 \exp \left( {-1.293 + \mu_{e^-} \over T_e} \right) C_3;
\end{equation}
\begin{equation}
\label{eq:e+rate}
\lambda_{e^+} \approx (1.578 \times 10^{-2} \, {\rm s}^{-1})
 \left({T_e \over m_e c^2} \right)^5
\exp \left( {-0.511 - \mu_{e^-} \over T_e} \right) C_4. 
\end{equation}
Here $C_3$ and $C_4$ are defined in the following way:
\begin{equation}
C_3 \approx 1+ (0.646 \, {\rm MeV})/T_e + (0.128 \, {\rm MeV}^2)/T_e^2 ;
\end{equation}
\begin{equation}
C_4 \approx 1 + (1.16 \, {\rm MeV})/T_e + (0.601  \, {\rm MeV}^2)/T_e^2 
+ (0.178  \, {\rm MeV}^3)/T_e^3 + (0.035  \, {\rm MeV}^4)/T_e^4.
\end{equation}
In these expressions, as everywhere in this paper, 
the total electron chemical potential
$\mu_{e^-}$ is defined as the kinetic chemical potential plus the rest mass of
the electron.
We have estimated the value of the Fermi integrals (see Fuller, Fowler, \&
Newman 1985 equations 5a-f)
assuming $ (m_p - m_n + \mu_{e^-}) / T_e  << 0$.  
Clearly, this assumption is only
valid when the electrons are not very degenerate.  The region we are studying
is far from the neutron star, and
at sufficiently high temperature and low density that this approximation is
reasonable.  These rates depend strongly on temperature, 
which causes their magnitude
to fall quickly in the outflowing material. 
Because of the many approximations involved in obtaining these
expressions, they may not faithfully
 represent the true rates to the accuracy necessary for 
nucleosynthesis calculations. 
Therefore, we use them only to illustrate their influence on the electron
fraction.  For example, convection or other multidimensional effects could
necessitate following the lepton capture rates in degenerate conditions
which would modify the evaluation of the rates in equations (21) and (22). 
The non-equilibrium calculation presented in section 
\ref{sec:noneq} includes a
more precise calculation of the electron and positron capture rates.  

The first question that we address concerns the relative magnitude of these
rates.  We can combine the approximate expressions for the rates into a ratio:
\begin{equation}
\label{eq:elrat}
{\lambda_{e^-} \over \lambda_{e^+}} \approx \exp{\left[ 
\left( -0.782 + 2 \mu_{e^-} \right)/T_e \right]} \left(C_3 \over C_4
\right);
\end{equation}
where we intend $\mu_{e^-}$ and $T_{e}$ to be expressed in units of MeV.
The reduction in the ratio caused by the threshold for 
the electron capture reaction is contained in the term $\exp(-0.782 \,
{\rm MeV}/T_e)$. 
The expression in equation (\ref{eq:elrat}) also demonstrates the effect of 
the chemical potential in reducing the positron capture rate.  In general, the
temperature in the region of interest 
is not high enough to neglect the additional multiplicative term,
$C_3 / C_4$, which comes from the evaluation of the phase space factors.
  In
Figure \ref{fig:ratios1} we show a plot of this ratio.  The curve 
$\lambda_{e^-} / \lambda_{e^+}$
is depicted starting (far left) in conditions where
the system is in weak equilibrium, at a temperature of about $T_e \approx 2
\, {\rm MeV}$. 
This corresponds to a time of $t_{pb} = 0.575 \, {\rm s}$ in the figure. 

In our calculated estimates of $\lambda_{e^-} / \lambda_{e^+}$
in Figure \ref{fig:ratios1} we have utilized an initial $Y_e$ from the weak 
equilibrium
condition, a time dependent density fit to the Mayle and Wilson supernova
computational results, and a constant value of the entropy per baryon set to
$s/k \approx 80$.  In an actual supernova we would not expect the entropy of 
the
outflow to be constant, nor would we expect $s/k = 80$ to obtain 
necessarily.  We
pick $s/k \approx 80$ and approximate the outflow as adiabatic for 
illustrative purposes
only.  In section 6, we consider other values of $s/k$.
Our time dependent density fit can be taken to represent
the density history of an outflowing Lagrangian mass zone during a limited
time period.  The functional
form we have adopted for this fit is:
\begin{equation}
\label{eq:fitdens}
\rho \approx \exp \left( -310.154 + 1811.43t - 3215.62 t^2 + 1839.84
t^3 \right) {\rm g \, cm^{-3}};
\end{equation}
where t is time post core bounce.
We have calculated all other thermodynamic variables, such as the temperature 
and chemical potential, from this density fit and assumed entropy.  
If the assumed entropy is decreased, then the effect of electron 
degeneracy will exhibit itself in the ratio of the rates.  
In this case, the ratio of the
electron capture rate to the positron capture rate  will be greater than
that shown in Figure \ref{fig:ratios1}, since the number density of 
electrons will be enhanced, while the
the number density of positrons will be suppressed.  The ratio 
$\lambda_{e^-} / \lambda_{e^+}$
decreases at lower
temperature (larger time), due to the increasing importance of the threshold 
for electron capture,  and also because of 
the increasing ratio of positron number 
density
to electron number density.  It can be seen from this figure that the ratio of
the electron capture rate  to the positron capture rate 
exhibits much more variation than the ratio
of the antineutrino capture rate to the neutrino capture rate on free
nucleons.

The ratio of the positron capture rate to the neutrino capture rate on free
nucleons can serve to illustrate an important point about the evolution of
conditions with radius and time in the outflowing material.
The ratios may be expressed as follows:
\begin{equation}
{\lambda_{e^+} \over \lambda_{\nu_e}} \approx 2.34 \,
\left({T_e \over {\rm MeV}}\right)^5 r_7^2 \left({
10^{51} {\rm ergs} \over L_{\nu_e} } \right) 
\left({{\rm MeV} \over T_{\nu_e}}\right) \exp \left[ 
\left( -m_e - \mu_{e^-} \right)/T_e \right] \left( C_4 \over C_1 \right);
\end{equation}
where the electron rest mass is $m_e \approx 0.511 \, {\rm MeV}$.
In Figure {\ref{fig:ratios2}} we plot $\lambda_{e^+} / \lambda_{\nu_e}$ as a
function of time.  It is evident from this figure that as a mass element
flows away from the neutron star, $\lambda_{e^+}$ drops off more quickly than
does $\lambda_{\nu_e}$. The drop in $\lambda_{e^+}$ is a result of the rapid
decrease in plasma temperature with radius, while the drop in 
$\lambda_{\nu_e}$ with radius simply reflects the relatively slower fall off
in the neutrino flux (the $1/r_7^2$ term in equation \ref{eq:nuapp}).
The decrease in the electron chemical potential $\mu_{e^-}$ with radius has a
relatively smaller effect in comparison to the fall in plasma temperature in
setting $\lambda_{e^+} / \lambda_{\nu_e}$. 

Since, as shown above, the electron and positron capture rates 
on free nucleons are small
relative to the corresponding
neutrino and antineutrino capture rates, the electron fraction
may be written as an expansion in the small parameter $\lambda_{e^+} /
\lambda_{\nu_e}$.  Employing this small parameter, we can expand the
expression for the electron fraction in equation (\ref{eq:soldiff}) to yield,
\begin{equation}
Y_e \approx Y_{e_0} \left \{ 1 + {\lambda_{e^+} \over \lambda_{\nu_e}} 
\left[ 1 -
(1 + \lambda_{e^-}/\lambda_{e^+}) / (1+ \lambda_{\nu_e} / 
\lambda_{\bar{\nu}_e})
\right] \right \} + {\cal O} \left[ \left({\lambda_{e^+} \over 
\lambda_{\nu_e}}\right)^2 \right].
\end{equation}
The first order term in this equation is sufficient to show the effect of
electron (positron) capture rates on the \lq\lq equilibrium\rq\rq\ $Y_e$.
  Since the lepton capture processes
are not necessarily in true chemical or steady state
equilibrium, by \lq\lq equilibrium\rq\rq\
 $Y_e$ here we mean the $Y_e$ that would
obtain if the system {\it were}
 in true equilibrium.  As the material flows outward in the region of
interest, the term $\lambda_{e^+} /
\lambda_{\nu_e} $ decreases rapidly. This term represents the competition
between increasing distance from the neutron star and decreasing 
plasma temperature.
  However, the term $\left[ 1 -
(1 + \lambda_{e^-}/\lambda_{e^+}) / (1+ \lambda_{\nu_e} / 
\lambda_{\bar{\nu}_e})
\right] $ actually shows a slight increase with radius or time.
This is caused by the decreasing ratio
$\lambda_{e^-} / \lambda_{e^+}.$  In the example discussed here 
the increase in the term 
$\left[ 1 -
(1 + \lambda_{e^-}/\lambda_{e^+}) / (1+ \lambda_{\nu_e} / 
\lambda_{\bar{\nu}_e})
\right] $
is overwhelmed by 
the decreasing positron capture rate in the leading $\lambda_{e^+} /
\lambda_{\nu_e}$ term, so that the change in the 
\lq\lq equilibrium\rq\rq\ electron fraction with radius is dominated by the
behavior of $\lambda_{e^+} / \lambda_{\nu_e}$.
 
Figure \ref{fig:yeepem} shows this \lq\lq equilibrium\rq\rq\ 
electron fraction.  The upper curve labeled \lq\lq equilibrium $Y_e$\rq\rq\
includes the effect of electron and positron
captures on free nucleons.  The lower curve in this figure labeled 
\lq\lq $Y_{e_0}$\rq\rq\, includes only neutrino captures on free nucleons. 
  At high temperature
 ($T_e \approx  2 \, {\rm MeV}$) corresponding to early times, 
the electron and positron captures make 
a considerable difference in the electron fraction (on the order of
 $\approx 10\%$).  
In fact, we have
slightly underestimated their effect by using the approximate expressions 
for the rates in equations (\ref{eq:e-rate}) and (\ref{eq:e+rate}).
However, it is clear that these processes
 become less important with time, as the \lq\lq
equilibrium\rq\rq\ electron fraction slowly asymptotes to the value of
$Y_{e_0}$.  In more electron 
degenerate conditions, the electron capture rate may be
greater than the positron capture rate, causing the \lq\lq
equilibrium\rq\rq\ electron fraction to lie below $Y_{e_0}$ at early times.
In this case,  the electron and 
positron capture processes change 
the \lq\lq equilibrium\rq\rq\ electron fraction by about
$2\%$ just before alpha particle formation.  We emphasize that this conclusion
depends on our particular Mayle \& Wilson outflow history.  Different models will
give disparate $Y_e$ values (the results of section 6 for different entropies
could 
be used to gauge how different $Y_e$ might be in other, more realistic models).

\section{The Alpha Effect}
\label{sec:alpha}

In this section we use the \lq\lq
equilibrium\rq\rq\ $Y_e$ to discuss qualitatively
 the behavior of the electron fraction during alpha
particle freeze-out.
The formation of alpha particles occurs while the outflowing plasma is 
still in nuclear
statistical equilibrium (NSE).  
Therefore, the mass fraction of alpha particles 
at a given density and temperature is well approximated
 by the Saha equation:
\begin{equation}
X(N,Z) = {G(N,Z) \over \rho N_A} A^{5/2} \left[ {2 \pi \hbar^2 N_A \over 
kT_e}
\right]^{3(A-1)/2} \left( {\rho N_A X_n \over A_n }\right)^N
\left( {\rho N_A X_p \over A_p }\right)^Z 2^{-A} \exp \left[ Q_n(N, Z) /kT_e
\right]
\end{equation}
\begin{equation}
\label{eq:sahaalpha}
X_\alpha \approx 
{3.256 \times 10^{-5} \rho_{10}^3 \over {(T_e / {\rm MeV})}^{9/2}} X_n^2
X_p^2 \exp(28.29 \,{\rm MeV} /T_e),
\end{equation}
where $X(Z,N)$ is the mass fraction of the nuclear species with $Z$ protons
and $N$ neutrons,
$G(Z,N)$ is the partition function, $\rho$ is density, $N_A$ is
Avagadros' number, $k$ is the Boltzmann constant, $Q_n(N,Z)$ is the nuclear
$Q$-value, and $A_n$ and $A_p$ are the neutron and proton atomic masses,
respectively.  In equation (\ref{eq:sahaalpha}),
$\rho_{10}$ is the density in units of $10^{10} {\rm g \, cm^{-3}}$.
In order to estimate the number density of alpha particles, we use a
plasma
temperature calculated from a given entropy and the fitted density function
(equation \ref{eq:fitdens}) as discussed in the last section.

To zero order in the ratio $\lambda_{e^+} / \lambda_{\nu_e}$, the electron
fraction in the presence of an alpha particle component
is given by (see equation \ref{eq:soldiff}),
\begin{equation}
\label{eq:eqyealpha}
Y_e \approx Y_{e_0} + X_\alpha (0.5 - Y_{e_0}).
\end{equation}
The magnitude of the \lq\lq alpha effect\rq\rq\ 
 on $Y_e$ is proportional to the number of alpha particles and also to
the difference between $Y_{e_0}$ and $1/2$.  If there are equal numbers of
neutrons and protons, then there is no change in the electron fraction due to
alpha particle formation.  However, if the neutron-to-proton ratio is not 
unity, then the effect of alpha particle formation
 will be to drive $Y_e$ closer to 1/2.  The expression in equation
(\ref{eq:eqyealpha})
 does not include the small effect of electron and positron capture
during this period.  These captures
 may be accounted for by including the terms to first
order in the ratio $\lambda_{e^+} / \lambda_{\nu_e}$:
\begin{eqnarray}
\label{eq:eqyealpha2}
 Y_e \approx && Y_{e_0} \nonumber \\
&&+ Y_{e_0}  {\lambda_{e^+} \over \lambda_{\nu_e}} \left[ 1 -
(1 + \lambda_{e^-}/\lambda_{e^+}) / (1+ \lambda_{\nu_e} / 
\lambda_{\bar{\nu}_e}) \right] \nonumber \\
&& + X_\alpha (0.5 - Y_{e_0}) \nonumber\\ 
&& +  Y_{e_0} {X_\alpha \over 2} \left({\lambda_{e^+} \over \lambda_{\nu_e}}
\right)
\left[{\lambda_{e^-} \over \lambda_{e^+}} - 1 - 
\left({\lambda_{\bar{\nu}_e} \over
\lambda_{\nu_en}} - 1 \right) 
\left( {\lambda_{e^-} / \lambda_{e^+} + 1 \over 
\lambda_{\bar{\nu}_e} /
\lambda_{\nu_e}+ 1}  \right) \right].
\end{eqnarray}
The second line in this expression is the correction to the 
\lq\lq equilibrium\rq\rq\ $Y_e$ from electron and positron
captures on free nucleons 
discussed in the last section.  The last line in equation
(\ref{eq:eqyealpha2}) contains
an additional correction term for the \lq\lq equilibrium\rq\rq\ $Y_e$ due to
the electron and positron capture processes on free nucleons when alpha
particles are present.
These additional terms give a more complete description of the \lq\lq
equilibrium\rq\rq\ $Y_e$ when alpha particles are present.  
Figure \ref{fig:yexa} shows the value of
$Y_{e_0}$ and the value of the \lq\lq equilibrium\rq\rq\ $Y_e$ as a function
of time for our example outflow trajectory.  
 At early time and high
temperature, the influence of the electrons and positrons 
can be seen by the increase in the  \lq\lq
equilibrium\rq\rq\ $Y_e$ over $Y_{e_0}$. The minimum in the 
\lq\lq equilibrium\rq\rq\ $Y_e$ curve occurs at the point where alpha 
particles begin to form.  The increase in \lq\lq equilibrium\rq\rq\ $Y_e$
as time increases subsequent to the minimum
is due to the \lq\lq alpha effect.\rq\rq\  
Since the true $Y_e$ is not in equilibrium at the time of alpha
particle formation, the increase in the true 
$Y_e$ will be much smaller than the increase 
in the  
\lq\lq equilibrium\rq\rq\ $Y_e$ shown in the figure.  The actual increase in
the true $Y_e$ depends on the rate of alpha particle formation.

Since the process of incorporating nucleons into alpha particles
 is clearly an important factor for 
determining the electron fraction, we wish to explore the relationship 
between
the rates of neutrino and antineutrino capture on free nucleons and the 
rate of
alpha particle formation.  Assuming a {\it slow} and 
constant rate of alpha particle
formation $ \dot{X_\alpha} = C_\alpha$,  and assuming constant neutrino 
and antineutrino
capture rates, we can integrate the differential equation for $Y_e$
(equation \ref{eq:soldiff}) to obtain:
\begin{equation}
\label{eq:alphatime}
Y_e \approx Y_{e_i} + 
{\lambda_{\nu_e} X_\alpha^2 \over C_\alpha} \left( 1/(2 Y_{e_0}) - 1\right).
\end{equation}
Here $ Y_{e_i}$ is the value of the electron fraction before alpha particle
formation begins.  Equation (\ref{eq:alphatime}) will be valid during the
epoch where the alpha particle abundance is changing and is accompanied by a
continuously changing electron fraction.  
The presence of the term
 $\lambda_{\nu_e} / C_\alpha$ in equation (\ref{eq:alphatime}) 
shows that there is a competition
between alpha particle formation and the neutrino capture on neutrons in
setting $Y_e$.  Either a faster
neutrino capture rate or a slower rate of alpha particle formation will 
result in an increased change in the electron fraction.  If the material is
neutron rich this change will be positive.
The last term in this
expression shows that there will be little effect 
from alpha particle formation when the electron fraction is
close to $Y_e=  0.5$.  

The change in the electron fraction due to alpha 
particle formation
may be estimated from equation (\ref{eq:alphatime}).  
At $t_{pb} \approx 0.5 {\rm s}$ the 
appropriate
quantities taken from our calculations give $\lambda_{\nu_e} /
C_\alpha \sim 5 \,  {\rm s}^{-1} / 50 \, {\rm s}^{-1} \sim 0.1$ and 
$( 1/(2 Y_{e_0}) - 1) \sim 0.3$.  Therefore, 
the change in the electron fraction as computed from equation 
(\ref{eq:alphatime}) will be
 $\sim 0.003$,
approximately 1\%.  The non-equilibrium calculation presented in section 
\ref{sec:noneq}
gives the results of a more exact and detailed treatment of these issues.
These results can be used to estimate how much different $Y_e$ would be for
different models of thermodynamic history and (anti)neutrino energy spectra.

\section{Non-Equilibrium Calculations}
\label{sec:noneq}

In this section we focus on the non-equilibrium aspect
 of the calculation of the
electron fraction.  In the previous sections we have focused on the \lq\lq
equilibrium\rq\rq\ $Y_e$, in order to show the general trends caused
by each of the variables which influence the electron fraction.  However, the
relative magnitudes and effects
 of these factors can only be determined definitively by a
full non-equilibrium calculation.

For this calculation, we employ
an NSE computer code to keep track of thermodynamic variables, weak
reaction rates and the electron fraction in a 
representative mass element of outflowing
material.  As input, the code utilizes the density fit from equation 
(\ref{eq:fitdens}), an assumed initial electron fraction and a constant
 entropy as before.
The outflow velocity of the
material is obtained by interpolating with cubic splines between 
the velocity given in a particular run of the Mayle and Wilson results
at different time slices.
Other thermodynamic variables, such as temperature and chemical potential are
calculated self-consistently from these parameters as outlined in section 
\ref{sec:nucap},
but now modified where appropriate to include the effects of heavy nuclei.
We calculate
 the relative abundances of free nucleons, alpha particles, and
heavy nuclei from the thermodynamic variables and the current value of the 
electron fraction.  The electron fraction is updated at each time step in our
calculation by following all the charged current weak
reaction rates.

During the periods when only free nucleons are present, the
electron fraction determines the values of $X_p$ and $X_n$.  After alpha
particles form, relative abundances are calculated with
the electron fraction and the nuclear
Saha equation for alpha particles, equation (17). 
During the periods when heavy nuclei are present, we
use the liquid drop model and the prescription of Fuller (1982), Bethe,
Brown, Applegate and Lattimer (1979; hereafter BBAL),  
Baym, Bethe, and Pethick (1971)
and Lamb et al. (1978)
 to calculate the neutron and 
proton chemical potentials.  In this model, the total 
energy of the nucleus with mass number A can be
approximated as,
\begin{equation}
W_N \approx W_{\rm bulk} + W_{\rm surf}A^{2/3} + W_{\rm coul}A^{5/3}.
\end{equation}
In this expression, $W_{\rm bulk}$ is the energy of bulk nuclear matter, 
while
 $W_{\rm surf}$ and $W_{\rm coul}$ are the coefficients of the
surface energy and coulomb energy terms, respectively. 
The values for these coefficients given
in BBAL and employed by Fuller (1982) are,
\begin{equation}
W_{\rm surf} \approx 290 (Z/A)^2 (1 - Z/A)^2;
\end{equation}
\begin{equation}
W_{\rm coul} \approx 0.75 (Z/A)^2 ( 1 - 0.236 \rho_{12}^{1/3} 
+ 0.00194 \rho_{12}),
\end{equation}
where $\rho_{12}$ is the density in units of $10^{12} {\rm g} \, 
{\rm cm^{-3}}$, and
Z is the number of protons in the nucleus. 
The value of the mean nuclear mass in NSE 
 is obtained by minimizing $W_N$, yielding the condition that the
nuclear surface energy should be 
twice the Coulomb energy, from which we derive,
\begin{equation}
 A \approx 194(1 - (Z/A))^2 (1- 0.236 \rho_{12}^{1/3})^{-1}.
\end{equation}
The difference in neutron and proton chemical potentials is given
by,
\begin{equation}
\hat{\mu} =
\mu_n - \mu_p \approx 250 \left[ 0.5 - (Z/A) \right] 
- W_{\rm surf} A^{-1/3} \left\lbrace{3
- 5( Z/A) \over (Z/A) \left[ 1 - (Z/A) \right]}\right\rbrace.
\end{equation}
The neutron chemical potential in this scheme is 
(Fuller 1982; BBAL),
\begin{equation}
\mu_n \approx -16 + 125 [ 0.5 - (Z/A)] -125 ( 0.5 - (Z/A))^2 - 
{ W_{\rm surf} \over A^{1/3} } {( 3 - 7 (Z/A) \over 2 [1 - (Z/A)]}.
\end{equation}
For a recent calculation and discussion of the nuclear equation of state see
Lattimer \& Swesty (1991).

The neutron chemical potential has the same value for the neutrons inside the
nucleus and the neutrons in the free neutron gas.
In the dilute Maxwell-Boltzman limit which obtains for free nucleons,
the mass fraction of free neutrons will be,
\begin{equation}
X_n \approx 79 {({T/ {\rm MeV}})^{3/2} \over \rho_{10}} \exp(\mu_n/T_e),
\end{equation}
while the corresponding mass fraction for free protons will be,
\begin{equation}
 X_p \approx X_n \exp(-\hat\mu / T_e). 
\end{equation}

The mass fractions of heavy nuclei are calculated by using the nuclear 
Saha
equation.  The partition functions and binding energies 
employed in this prescription were
taken from Woosley, Fowler, Holmes and Zimmerman (1978).  Only heavy nuclei
which contribute more than a few percent to the total 
heavy nucleus mass fraction
at a given time are retained,
and the abundance distribution is normalized to insure that
\begin{equation}
X_p + X_n + X_\alpha + \sum_h X_h = 1. 
\end{equation}

We allow the electron fraction to change
by calculating weak capture rates on free
nucleons and heavy nuclei.  In the case of (anti)neutrino capture on free
nucleons, the neutrino
distribution functions (smoothed with time)
 are employed as discussed in section \ref{sec:nucap}.  
The electron and positron capture
rates are calculated for arbitrary degeneracy and without
making any approximations in the lepton kinematics:
\begin{equation}
\lambda_{e^-} \approx (6.295 \times 10^{-4} \, {\rm s^{-1}}) 
\left(T_e \over m_e\right)^5
\int_{E_{TH}^-}^\infty { (x + Q_{e^-}/T_e)^2 x (x^2 - (m_e / T_e)^2)^{1/2} 
\over \exp(x - \mu_{e^-}/T_e) + 1} dx;
\end{equation}
\begin{equation}
\lambda_{e^+} \approx (6.295 \times 10^{-4} \, {\rm s^{-1}}) 
\left(T_e \over m_e\right)^5
\int_{E_{TH}^+}^\infty { (x + Q_{e^+}/T_e)^2 x (x^2 - (m_e/T_e)^2)^{1/2} 
\over \exp(x - \mu_{e^+} / T_e) +1} dx.
\end{equation}
Here the Q-values and the energy thresholds are
$Q_{e^+} \approx 1.293 \, {\rm MeV}$ and $E_{TH}^+ = m_e$ for 
positron capture on neutrons and 
$Q_{e^-} \approx -1.293 \, {\rm MeV}$ and $E_{TH}^- = m_n - m_p - m_e$
  for electron capture on protons.  In electromagnetic equilibrium, 
the sum of the 
total (kinetic plus rest mass) electron and positron chemical
potentials is zero, implying that
\begin{equation} 
 \mu_{e^+} = - \mu_{e^-}.
\end{equation}

We have included neutrino and anti-neutrino captures on
heavy nulcei, calculated by the prescription given in Fuller \& Meyer (1995) 
and McLaughlin \& Fuller (1995).  In these calculations, we employed 
Fermi-Dirac distribution functions for the
neutrinos and antineutrinos normalized by the appropriate average energy
 and luminosity.  The neutrino and antineutrino energy
luminosities and average energies were taken from the Wilson and
Mayle output at the relevant time slices.  The average neutrino and
antineutrino energies were fit with
straight lines, while the corresponding 
values for the luminosities were obtained using
cubic spline interpolation. 

We start our calculations in conditions where weak equilibrium obtains, so
that we can employ the solution to equation (\ref{eq:yeeqel}) 
as an initial electron fraction.  We hold
the entropy constant throughout our calculation, since we are 
far from the gain radius when the calculation begins.  
The adiabatic outflow
approximation is employed here in keeping with our spirit of 
discerning the basic
physics important for $Y_e$.  We would expect the entropy of an actual 
outflowing
mass element to rise moderately through the regime of weak freeze-out 
(cf. Qian \&
Woosley 1996). 
The resulting 
\lq\lq non-equilibrium\rq\rq\ electron fraction, for three
different constant entropy trajectories
 is shown in Figure \ref{fig:noneq}.  The resulting electron fraction
for all the curves follows a downward
trend at the earliest times.  This trend
 corresponds to the effect of the decreasing
importance of the electron and positron capture rates, as discussed in 
section \ref{sec:nucap}.
The curve with the highest entropy begins with 
highest electron temperature, since
the density history
 is the same for all of these curves.  This also explains the increase
in the electron fraction with increasing entropy, since the plasma
temperature and, hence, the effect of electron and positron capture will be
larger at higher entropy.  The dip in these curves
occurs when alpha particles begin to form.  This is a smaller version of
the effect in the \lq\lq equilibrium\rq\rq\ case which was 
discussed in section \ref{sec:alpha}.  It can be seen that the
smallest \lq\lq alpha effect\rq\rq\ occurs for the case where the
electron fraction is closest to 1/2.  

In this graph, heavy nuclei form starting at an epoch between,
$ t_{pb}  \approx 0.64 \, {\rm s} $ and $t_{pb} \approx 0.66 \, {\rm s}$.  
After the heavy nuclei form, the system contains mostly
heavy nuclei with a few percent of the mass fraction in protons. 
 Antineutrinos are capturing on protons, and this
 tends to decrease the electron fraction. 
However, counteracting this effect are neutrino captures on heavy nuclei, 
which tend to increase the electron fraction.  The result is the almost 
flat curves seen at
late time in Figure \ref{fig:noneq}.  Figure \ref{fig:heavy}
 further demonstrates the effect of neutrino
and antineutrino capture on heavy nuclei for a trajectory with entropy $s/k
\approx 40$.  This figure shows the 
$Y_e$ which is obtained, 
when captures on heavy nuclei are included (upper curve) and not
included (lower curve) in the calculation.  The captures on heavy 
nuclei represent a relatively small effect, which does not affect the 
resulting electron fraction by more than 1\%.

We find that the value of the electron fraction is close to $Y_{e_0}$
throughout 
the time period where the system is falling out of weak equilibrium. 
Although, the formation of alpha particles, neutrino and antineutrino 
capture on
heavy nuclei, and electron (positron) capture on free nucleons can have 
significant leverage
on the final electron fraction, the ratio of the antineutrino to neutrino
capture rate on free nucleons at the time of freeze out from weak 
equilibrium has the most influence.  However, the relative leverage on $Y_e$
of the effects considered here
can vary with different outflow conditions.

\section{ Conclusion}
\label{sec:conclusion}

In this paper we have given an in depth treatment and analysis of the
evolution of the electron fraction in neutrino-heated supernova outflow,
including detailed treatment of the effects of nuclear composition changes. 
Our study has concentrated on the time $t_{pb} \lesssim 1 \, {\rm s}.$
The evolution of $Y_e$ is this epoch may be quite important for models of the
light $p$-Process and the neutron number $N \approx 50$ $r$-Process nuclei
(Hoffman et al. 1996a; Fuller \& Meyer 1995).

We have employed fits to the detailed neutrino and antineutrino energy
spectra from the Mayle and Wilson supernova calculations.  We find that these
detailed spectra are necessary for computations of weak rates to the level of
accuracy in $Y_e$ which may be required for nucleosynthesis considerations. 
However, we find that hydrodynamic wave-induced
 fluctuations in the ratios of neutrino and
antineutrino spectral parameters with time produce significant excursions
in $Y_e$.  We find that the rates of electron and positron capture on free
nucleons can also be important for computing the evolution of the electron
fraction.  The charged current weak rates freeze out from equilibrium at a
time when the electron and positron capture rates may still have some
influence on $Y_e$.  During this time period, the effect of these rates is to
increase the electron fraction.  
We have given detailed calculations of the \lq\lq alpha effect\rq\rq\ - the
increase in the electron fraction caused by a changing alpha particle mass
fraction.  Our results indicate that the alpha effect can play a very
significant role in setting $Y_e$.

We have employed numerical calculations of nuclear composition changes in
nuclear statistical equilibrium, coupled with fits to density and velocity of
outflow histories from the Mayle and Wilson results, to compute the evolution
of $Y_e$.  These calculations explicitly include all charged current weak
interaction processes, including neutrino and antineutrino capture on heavy
nuclei.  We follow the evolution of $Y_e$ through the epoch of weak
equilibrium freeze out.  These calculations show that the combination of
neutrino and antineutrino capture on
heavy nuclei and antineutrino capture on free protons
tend to keep $Y_e$ constant.  The results presented in this
paper are meant to illustrate several different variables and processes which
can alter the electron fraction in post core bounce supernova outflow.  
Clearly,
more sophisticated models of neutrino transport and hydrodynamic outflow 
than those
employed  here would be necessary to actually {\it predict} $Y_e$ in a 
reliable
fashion.  We believe, however, that the effects described here will 
always play the
major role in setting $Y_e$.

\acknowledgements
We wish to acknowledge useful discussions with Y.-Z. Qian, B. S. Meyer,
S. E. Woosley, and R. D. Hoffman.
We also wish to thank R. Mayle for the use of output from the Mayle \& 
Wilson supernova code.
This work was support by NSF grant PHY-9503384 and a NASA theory grant at 
UCSD,
while JRW was supported by NSF grant PHY-9401636.

\newpage
\begin{table}
\begin{center}
\begin{tabular}{lllc} \hline\hline
\multicolumn{1}{c}{method} &
\multicolumn{1}{c}{$\lambda_{\nu_e} \, ({\rm s}^{-1})$} &
\multicolumn{1}{c}{$\lambda_{\bar{\nu}_e} \, ({\rm s}^{-1})$} &
\multicolumn{1}{c}{$Y_{e_0}$} \\ \hline
blackbody spectra&$39.10$&$45.86$&$0.4602$\\
cubic spline interpolation&$35.61$&$38.96$&$0.4775$\\
pieces of blackbody spectra&$35.54$&$38.69$&$0.4787$\\ \hline \hline    
\end{tabular}
\caption
[Comparison of (anti)neutrino spectra.]
{\label{tab:nccc}
Comparison of effect of (anti)neutrino spectra.}
\end{center}
\end{table}

\newpage
\figcaption[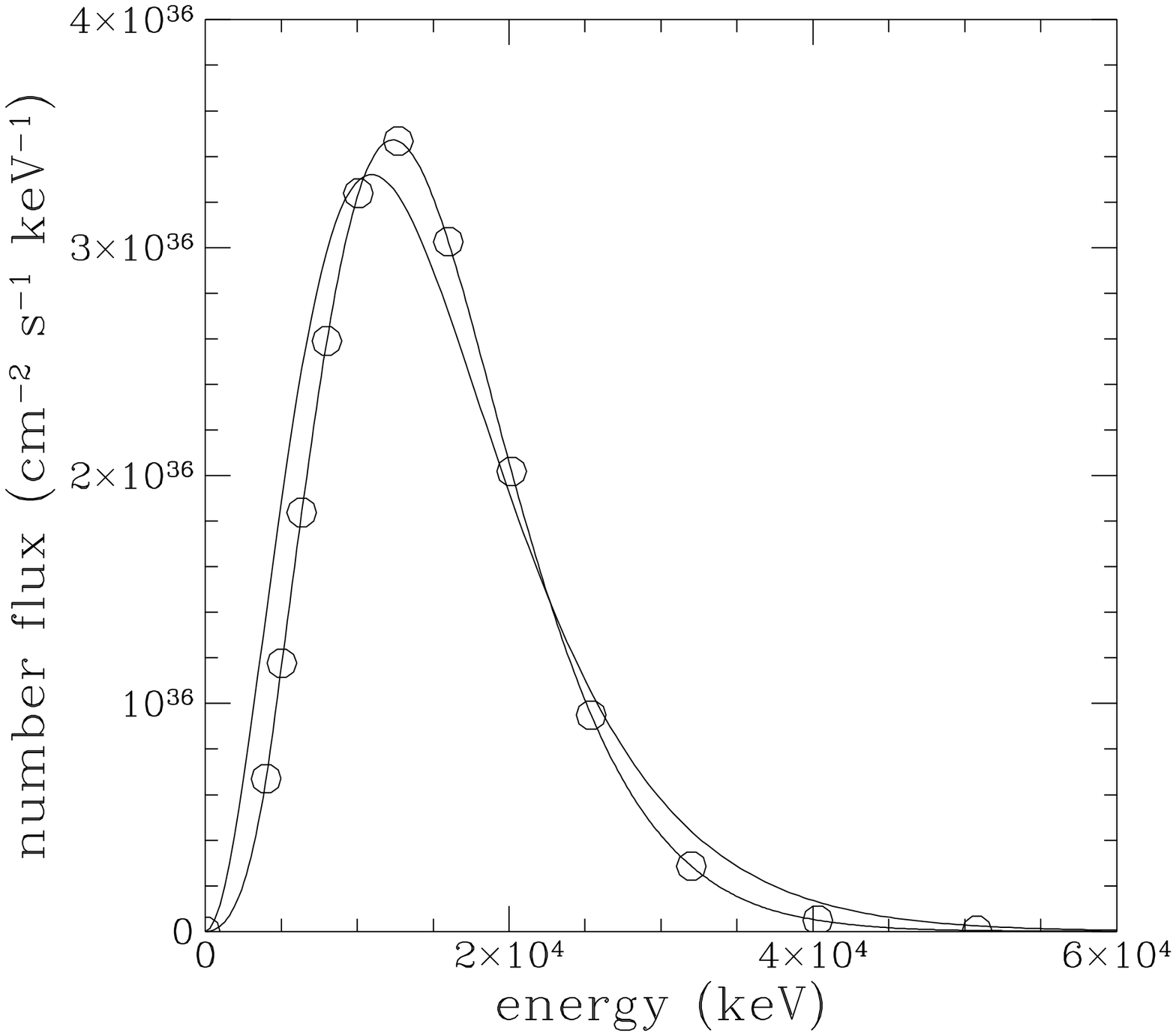]
{ \label{fig:compblack}
Comparison of the Mayle \& Wilson numerical transport calculation-derived
 neutrino energy spectrum with a zero chemical potential blackbody
spectrum.  Circles indicate the data
 points taken from the Wilson and
Mayle code.  The curve through these data 
points has been fit using cubic spline
interpolation.  The curve without circles 
shows a blackbody spectrum at a temperature of
$T_\nu = 3.15 \, {\rm MeV}$.}
%
\figcaption[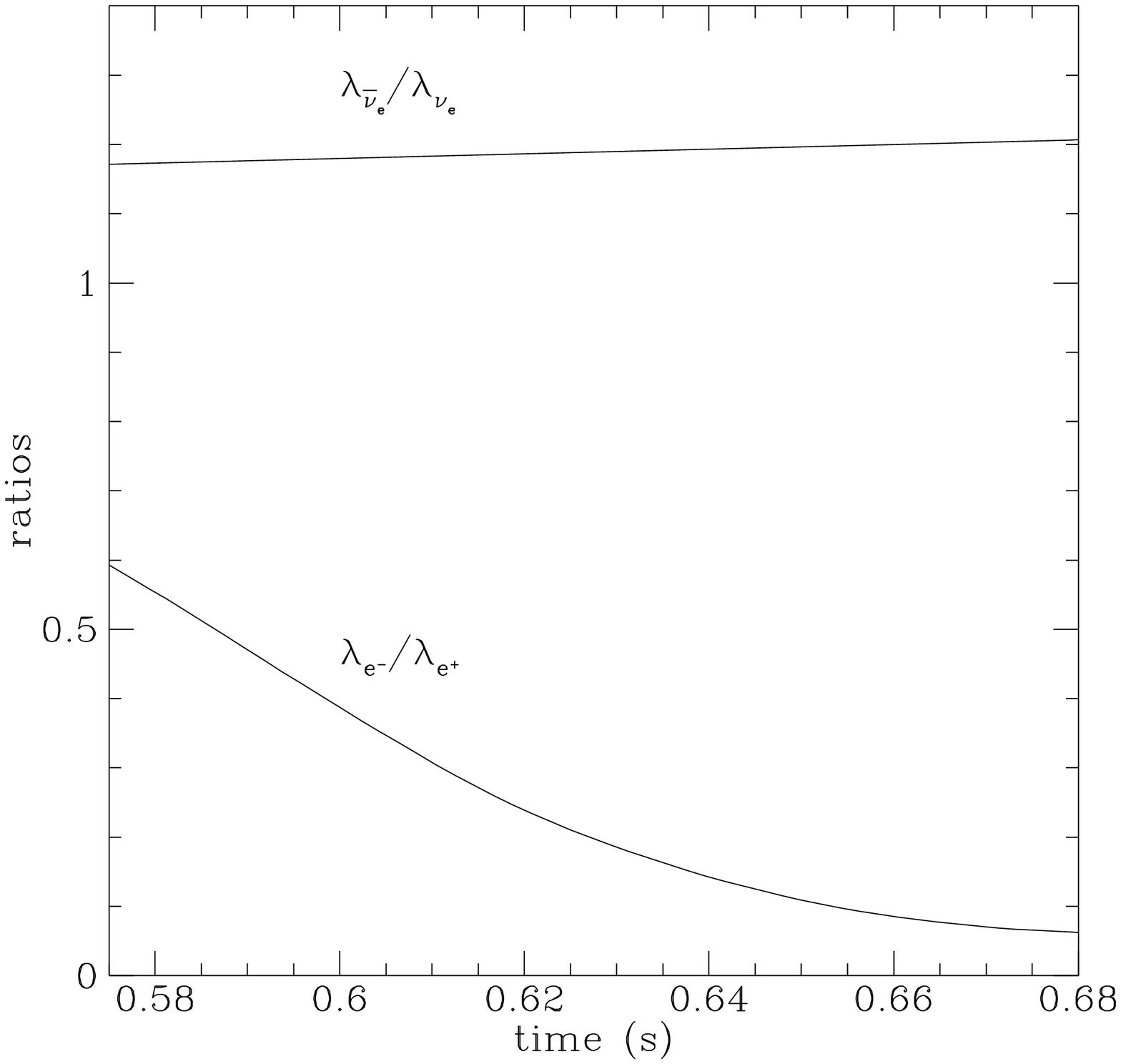]
{\label{fig:ratios1}
The upper curve is the ratio of the neutrino capture rate  on free neutrons
to the
antineutrino capture rate on free protons plotted against time.  
This curve has been smoothed in order
to average out fluctuations in the neutrino spectra. 
The lower curve is the ratio of the
electron capture rate on free protons
to the positron capture rate on free neutrons 
plotted against time, in a mass
element moving away from the neutron star.}
%
\figcaption[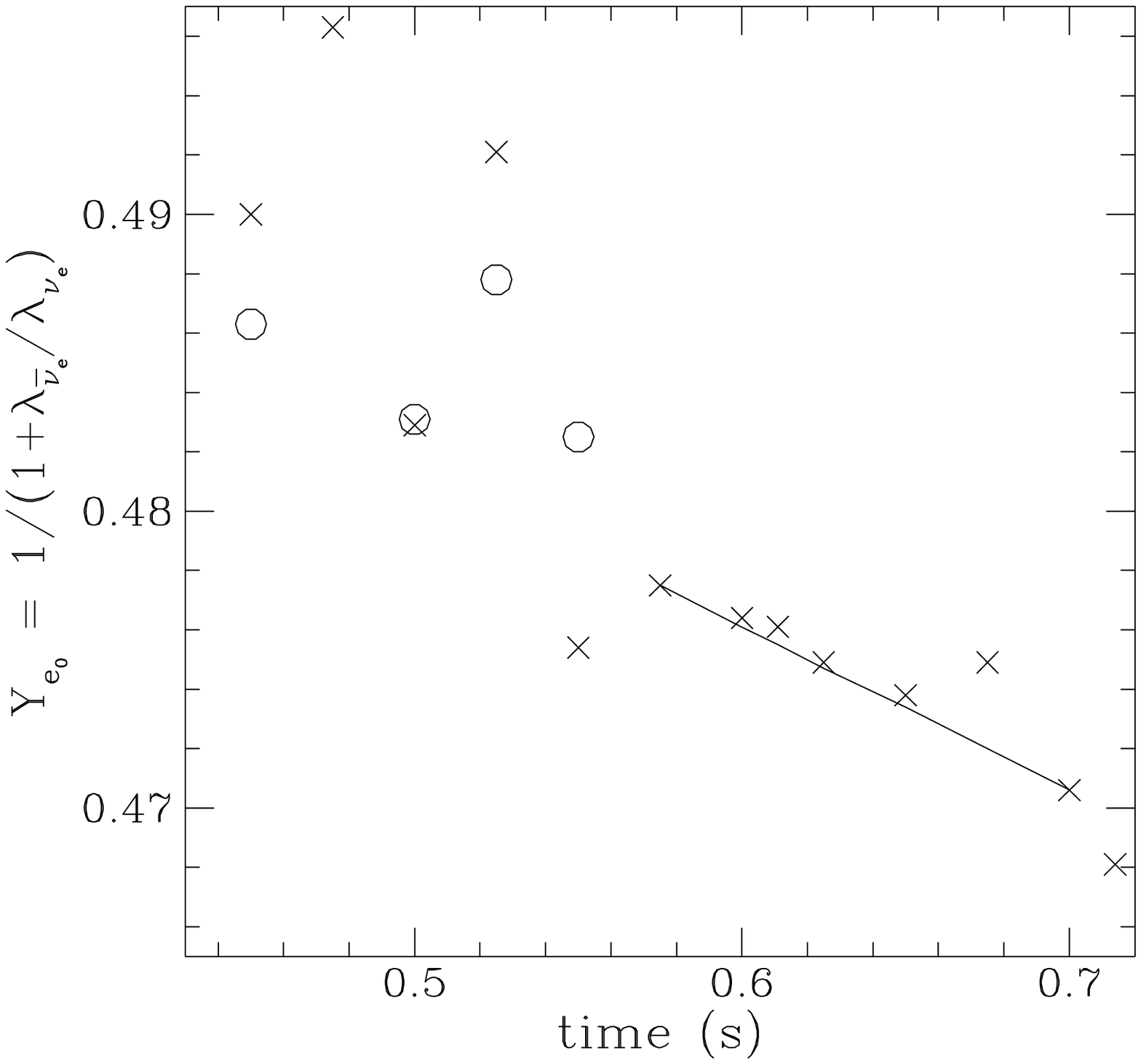]
{\label{fig:ye0vst}
Plot of the quantity $Y_{e_0}$ against time.  The crosses
correspond to points calculated from neutrino spectral data taken from
the Wilson and Mayle supernova code (crosses).  The same quantity derived
similarly from a different run of this code 
is also shown (circles).  The line
shows the function used in this paper to derive 
the quantity $Y_{e_0}$. }
%
\figcaption[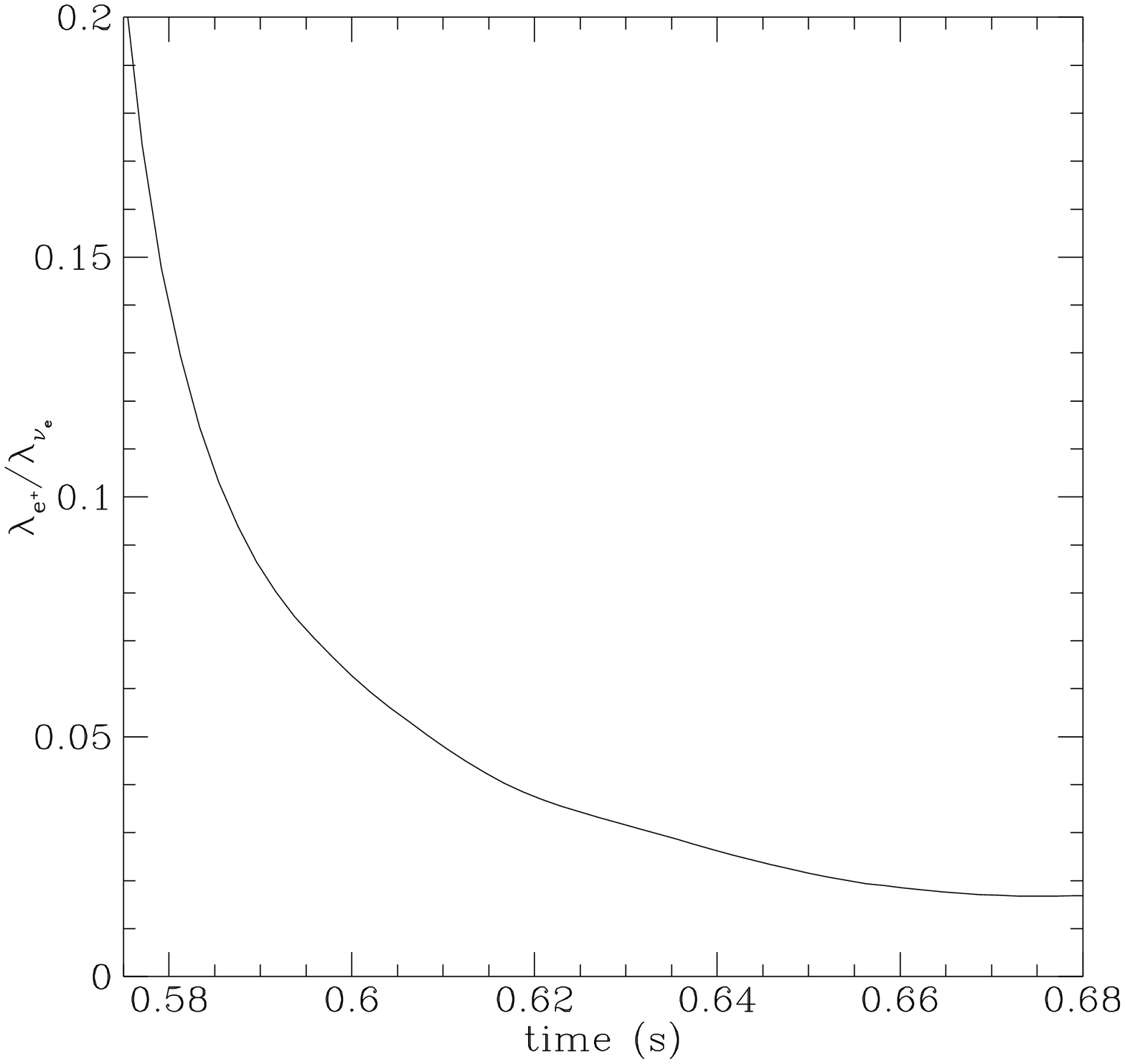]
{\label{fig:ratios2}
This figure shows the ratio of the positron capture rate on free
neutrons to the neutrino capture rate on free neutrons against time for an
outflowing mass element. }
%
\figcaption[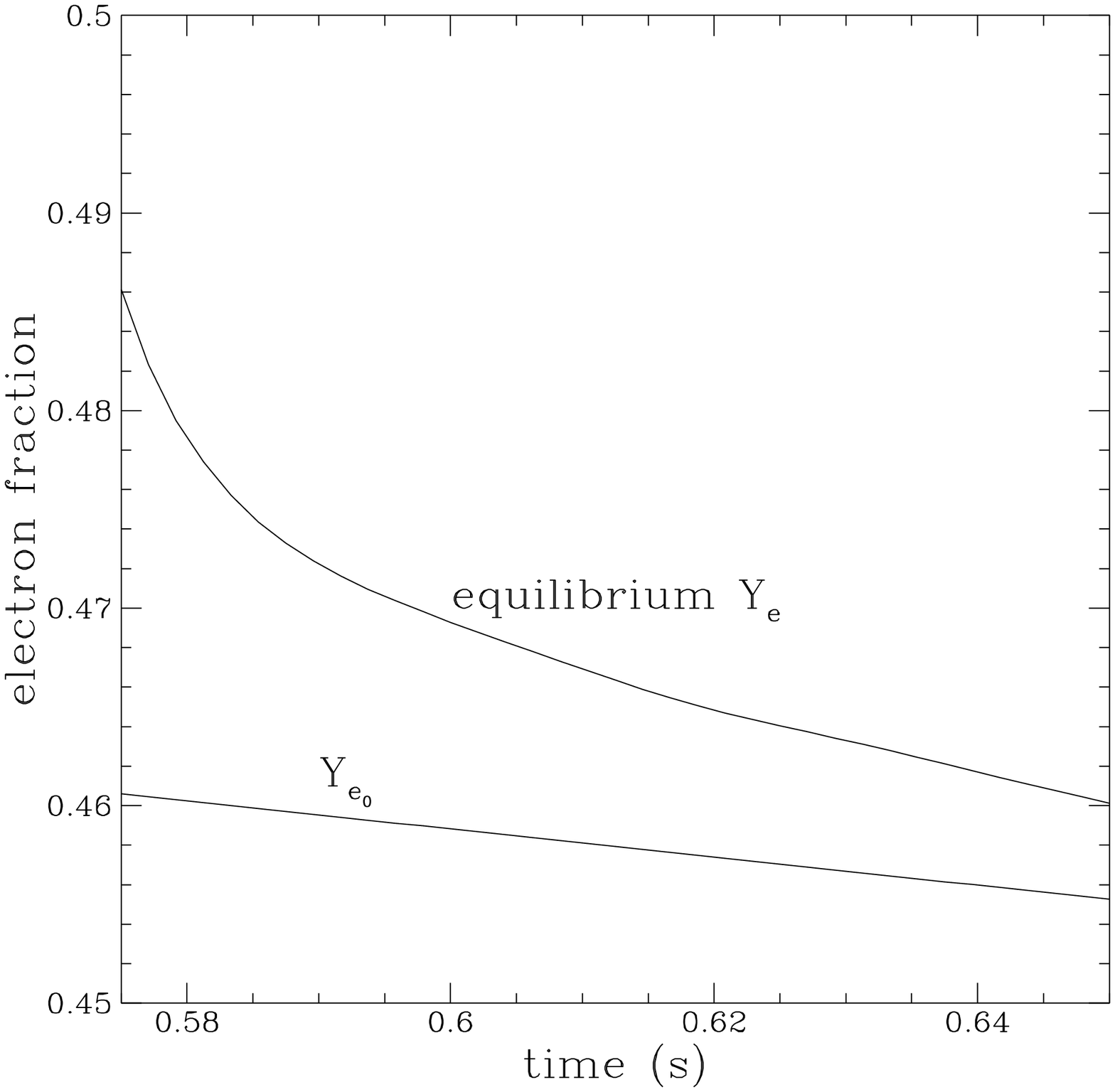]
{\label{fig:yeepem}
This figure shows the value that the electron fraction of an outflowing mass
element would take if the system were in weak equilibrium.
  The lower curve takes into account 
only neutrino and antineutrino captures on free nucleons. 
The upper curve also takes into account electron and positron
captures on free nucleons.  
In calculating these curves it has been assumed that the nucleons 
are
free at all times and have not been incorporated into nuclei. }
%
\figcaption[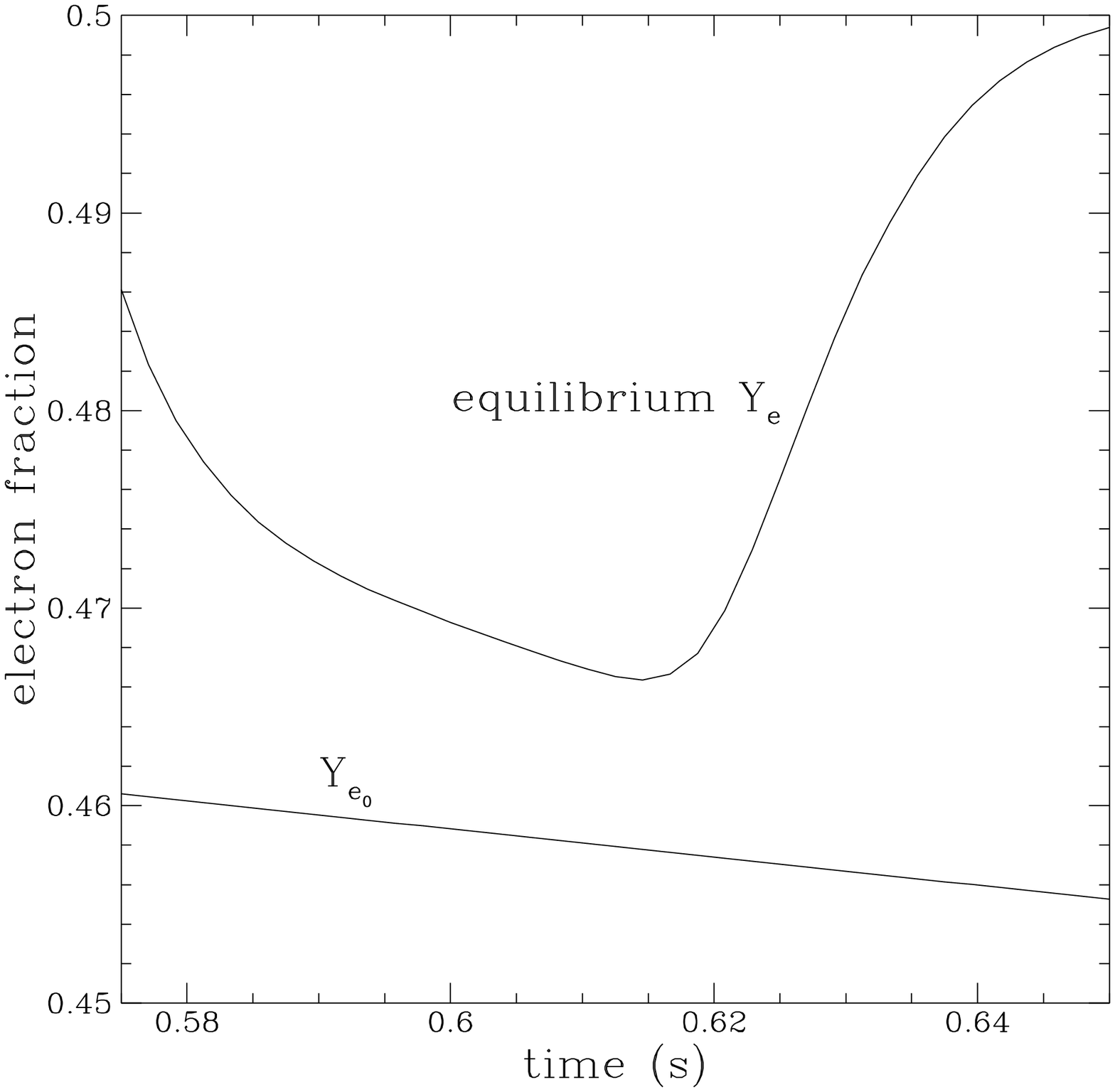]
{\label{fig:yexa}
 This figure shows the value that the electron fraction would take
if the system were in weak equilibrium.  The lower curve takes into account 
only neutrino and
antineutrino captures on free nucleons, as in Figure \ref{fig:ratios2}. 
The upper curve takes into account 
electron and positron captures on free nucleons, 
as well as the incorporation of free nucleons into alpha particles. }
%
\figcaption[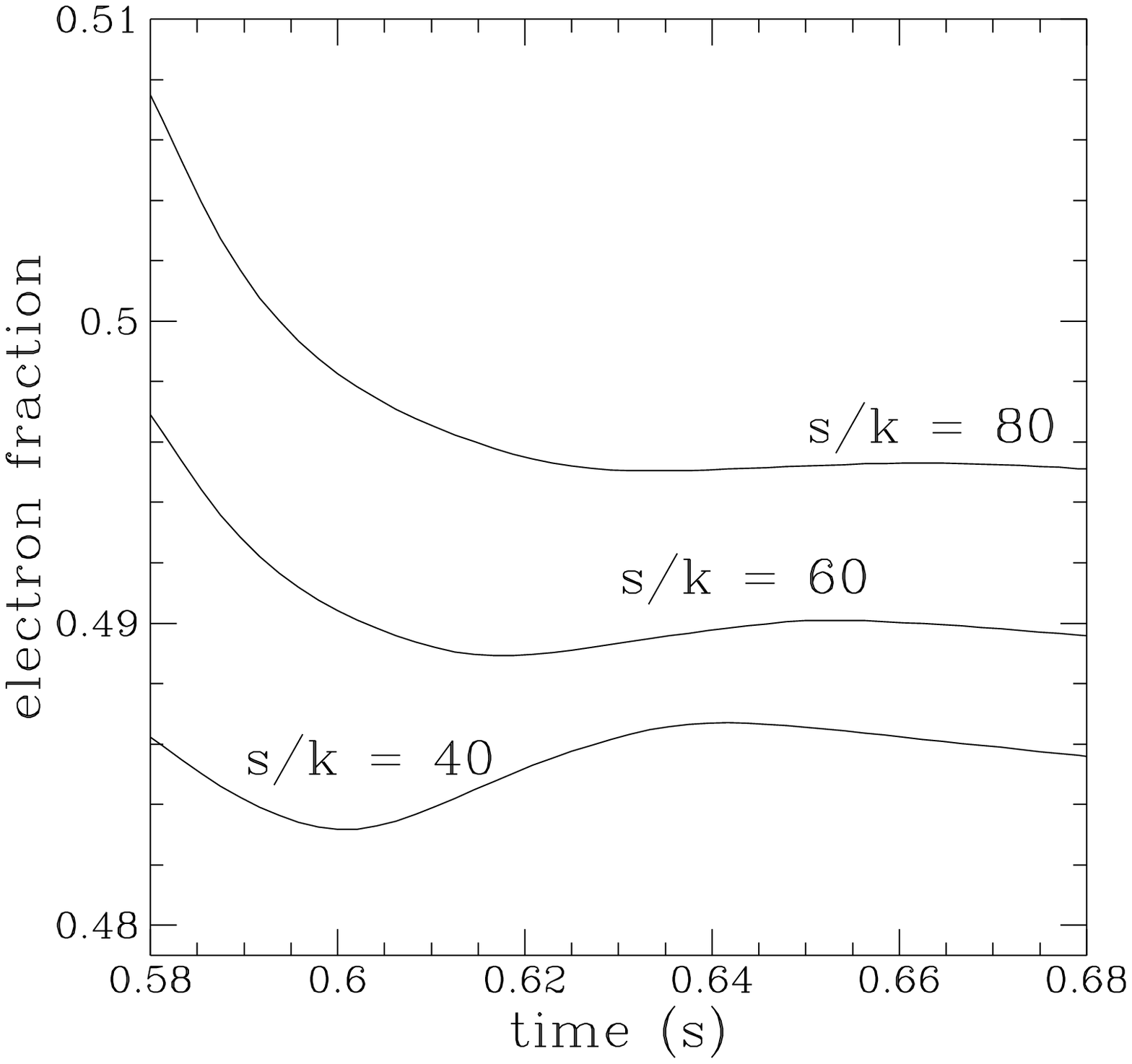]
{\label{fig:noneq}
This figure shows the value of the electron fraction derived from a
nonequilibrium calculation.  The entropy was held constant as the mass element
flowed away from the neutron star.  Three different curves are shown, each 
with a different value for the entropy, as indicated.}
%
\figcaption[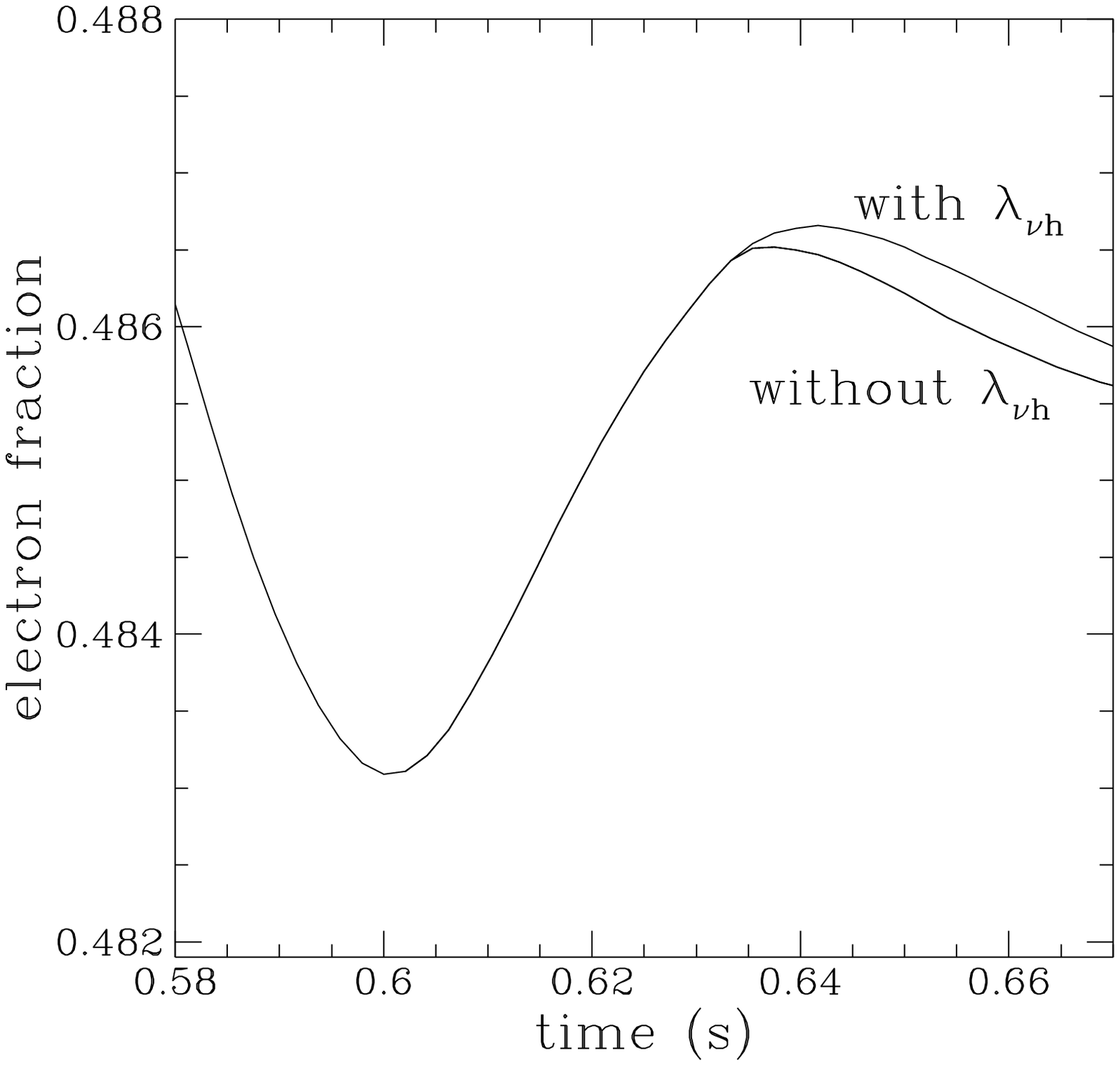]
{\label{fig:heavy}
Results of a nonequilibrium calculation for the electron
fraction are shown in which neutrino and antineutrino captures
on heavy nuclei are included 
(upper curve) and not included (lower curve).  The
electron fraction is plotted against time post core bounce.}

\end{document}